\documentstyle[amsmath,amssymb,graphicx,epic]{article}

\def\be{\begin{eqnarray}}
\def\ee{\end{eqnarray}}
\def\nn{\nonumber}

\def\Tr{{\rm Tr}\,}
\def\Pr{{\rm{\bf Pr}}}

\def\l[{\phantom.\! [}
\newcommand{\matr}[2]{\begin{array}{#1}#2\end{array}}
\renewcommand{\atop}[2]{\matr{c}{#1\\#2}}
\newcommand{\wt}{\widetilde}
\newcommand{\bld}{\textbf}

\newcommand{\brc}[1]{\left(#1\right)}

\makeatletter
\newcommand{\T}{
\@ifnextchar^{
    T   
}{
    T   
}}
\newcommand{\at}{
\@ifnextchar^{
    a
}{
    a
}}

\makeatother

\hoffset= - 1.0in         

\title{{\bf Differential hierarchy
and additional grading of knot polynomials } \vspace{.2cm}}
\author{{\bf S.Arthamonov}\thanks{{\small
{\it ITEP, Moscow, Russia}}; artamonov@itep.ru}, {\bf A.Mironov}\footnote{ {\small {\it
Lebedev Physics Institute} and {\it ITEP, Moscow, Russia}};
mironov@itep.ru; mironov@lpi.ru}, {\bf A.Morozov}\thanks{{\small
{\it ITEP, Moscow, Russia}}; morozov@itep.ru}}

\begin{document}
 \maketitle

\vspace{-6.5cm}

\begin{center}
\hfill FIAN/TD-09/13\\
\hfill ITEP/TH-20/13\\
\end{center}

\vspace{5cm}

\centerline{ABSTRACT}

\bigskip

{\footnotesize Colored knot polynomials possess a peculiar $Z$-expansion in certain
combinations of differentials, which depends on the representation.
The coefficients of this expansion are functions of the three variables
$A,q,t$ and can be considered as new distinguished coordinates on the space
of knot polynomials, analogous to the coefficients of alternative
character expansion.
These new variables are decomposed in an especially simple
way, when the representation is embedded into a product of the
fundamental ones.
The fourth grading recently proposed in \cite{GGS},
seems to be just a simple redefinition of these new coordinates,
elegant but in no way distinguished.
If so, it does not provide any new independent knot invariants,
instead it can be considered as one more testimony
of the hidden differential hierarchy (Z-expansion) structure behind
the knot polynomials.}

\section{Introduction}

Knot polynomials \cite{knopo} are observables (Wilson loop averages)
in one of the simplest Yang-Mills models:
Chern-Simons theory \cite{CS,Wit} in the simply connected $3d$
Euclidean space-time $M_3=R^3$ or $M_3=S^3$ (in more complicated spaces
these are even more interesting non-polynomial functions).
Studying these quantities and the rich set of relations between
them is a crucial step towards understanding the general structure
of Yang-Mills, gravity and more general string models.
Since the theory is topological, the Wilson loop averages\footnote{
In this paper we actually consider the {\it reduced} knot polynomials,
divided by their unknot counterparts,hence the $\sim$ sign
instead of the equality.}
\be
H_R^{\cal K}\ \sim\ \left<\Tr_{\!R\ } P\exp\oint_{\cal K} {\cal A}\right>
\ee
are relatively simple: they actually do not depend on
the geometry of the line ${\cal K}\subset M_3$, only on its
topological (linking) class, i.e. ${\cal K}$ can be considered as
a {\it knot} or a {\it link} (if it consists of several disconnected lines).
It also depends on representation $R$, on the coupling constant
$q=\exp\left(\frac{2\pi i}{k+N}\right)$ and on the gauge group
$G$, for which we take $G=SU(N)$; thus, irreducible representations $R$
will be finite-dimensional and labeled by the Young diagrams.
(generalizations to other Lie algebras, compact and  non-compact,
are rather straightforward).
As already mentioned,  for the simply connected $M_3$ this $H_R$
turns out to be a polynomial (called HOMFLY or HOMFLY-PT polynomial
\cite{knopo}) of the non-perturbative variables
$q$ and $A=q^N$, where $\log A$ is  the 't Hooft coupling constant,
which remains finite in the planar limit of Yang-Mills theory.

Not only it is a polynomial, all the coefficients turn out to be
integers what implies an additional  homological structure behind
the scene: the moduli of these integer coefficients count something
like (BPS) states of some hidden topological theory \cite{GoVa}.
Indeed, $H_R$ can be represented as the Euler polynomial of the
Khovanov-Rozansky complex \cite{Kh,KhR}, describing the flips
between different resolutions of the knot diagram induced by the
cut-and-join operators (see \cite{MD} for a recent review). The
corresponding Poincare polynomial has all the coefficients positive,
but depends on one extra parameter (grading) $T$, the original
polynomial \be H_R^{\cal K}(A|q) = \left.P_R^{\cal
K}(A|q,T)\right|_{T=-1} \ee is obtained from this "superpolynomial"
\cite{DGR} at $T=-1$.

For the HOMFLY polynomial $H_R$, there is now a very effective
method to evaluate them through a sum over paths in the representation
tree \cite{AnoMMMtree}.
It is based on
\begin{itemize}
\item
the quantum-${\cal R}$-matrix representation \cite{TR}
of knot polynomials in the temporal gauge $A_0=0$
(where the propagator gets ultralocal, $\delta^{(2)}(\vec x)\,{\rm sign}(t)$),
\item
braid representation of the knot diagrams ($2d$ projections
of ${\cal K}$ (naturally appearing in the temporal gauge)
and
\item
the character decomposition  \cite{DMMSS,MMMkn1}
(see also \cite{AS} and \cite{Che}),
which reduces the story to the much simpler $\hat{\cal R}$-matrices,
acting in the space of intertwining operators.
\end{itemize}

This method generalizes, conceptually and technically, the well-known
approach based on the skein relations in the fundamental representation
(where the $\hat{\cal R}$-matrix has just two eigenvalues and satisfies the equation
$({\cal} R-q)(q{\cal R}+1)=0$) and on the cabling approach, reducing
evaluation of colored knot polynomials to that of the fundamental ones for the cabled knots
(in modern terms it is sufficient to restrict the paths in the representation
tree to those passing through the vertex $R$).
It is technically very effective and allows one to calculate the colored HOMFLY
polynomials in rather sophisticated examples.
However, it is still not very clear, how this method is generalized
to superpolynomials.

Until recently, we actually knew some examples of superpolynomials
from two sources: from an artful analysis of particular cases in
\cite{DGR} and subsequent papers \cite{sp,AS,GS}, and from a
systematic {\it evolution method} of \cite{DMMSS,evo}, which,
however, needs some simple particular cases as the input. The most
spectacular application of evolution method is to the torus knots,
where, based on the original analysis of \cite{AS,DMMSS} and on a
specifically torus approaches of \cite{Gor}, a full solution is
suggested in terms of the affine Hecke algebras \cite{Che,Negut}.
However, it is unclear to what extent the evolution method is
systematically applied in this suggestion, and, most important, the
answers for the colored would-be-superpolynomials are not always
positive what can imply that they still describe some non-trivial
deformation of the Euler polynomial rather than the Poincare one. If
so, this implies the existence of new non-trivial deformations
(extra gradings), at least for the torus knots.

Fortunately, today we possess another powerful method to systematically
evaluate the superpolynomials: that of the $Z$-expansion \cite{IMMMfe}
(a far-going development of the original proposal in \cite{DGR}).
It is still underdeveloped, but it already allowed to calculate the
superpolynomials in all (anti)symmetric representations for
series of 2-bridge knots (twist and torus knots in \cite{inds} and
\cite{FGSto} respectively, where it is nicely consistent with
the evolution method, which allows to do even more \cite{evo})
and slowly proceed to less trivial representations \cite{AnoMMM21,GGS}.
But, most important, it provides a very different description of
knot polynomials and also sheds some new light on the problem
of extra gradings.
In particular, the recently suggested fourth grading of \cite{GGS}
seems to be one-of-many possible new gradings;
this view can help to understand that they all are still of lower
value then the original three variables $(A,q,T)$.
At the same time, the new superpolynomial examples found in \cite{GGS}
allow us to move further with understanding of the $Z$-expansion
approach.
Perhaps, a more adequate name would be "differential hierarchy"
referring to Khovanov-DGR "differentials" \cite{Kh,DGR}.
As to $"Z"$, it refers to significance of cases where the differentials
enter in pairs named $Z$-factors in \cite{IMMMfe}
(though this is literally so only for rectangular Young diagrams).
In what follows we use both names on equal footing.

\section{The idea of $Z$-expansion
}

Modern QFT approach to correlators in general, and to the Wilson loop averages in Chern-Simons theory in particular,
is to associate with any knot ${\cal K}$ and representation $R$
an element of an infinite-dimensional vector space
\be\label{off}
X_R^{\cal K}\ =\ \sum_I x_R^{\cal K}[I]\cdot e[I]
\ee
where $\{x_R^{\cal K}[I]\}$ is a set of coefficients of expansion in a fixed basis $e[I]$,
which does not depend on the knot and representation.
Then, the correlation function can be considered as a pairing of this element
and a point in the dual space which is called vacuum:
\be\label{on}
\hbox{correlation function}=\    <vac\,|\,X^{\cal K}_R>
\ee
In \cite{DMMSS,MMMkn1,MMMkn2} the elements like (\ref{off}) corresponding to the Wilson loop averages in Chern-Simons theory
and called {\it extended} or {\it off-shell}
knot polynomials were introduced in the space of symmetric functions with the bases of
characters, the Schur functions and the MacDonald polynomials of (infinitely many)
time-variables $\{p_k\}$. The pairing in that case is equivalent to
the reduction of the basis to the topological locus,
\be
p_k^*=\frac{A^k-A^{-k}}{t^k-t^{-k}},\ \ \ \ \ \ M_R(p^*)=M_R^*
\ee
(for  $\ A=t^N\ $  and $\ t=q\ $ these $\ M_R^*\ $ are just the ordinary quantum dimensions),
and the off-shell $X_R$ reduces to the {\it on-shell} HOMFLY polynomial $H_R$
(and the superpolynomial $P_R$) of the knot. Here $q$ and $t$ are the parameters of the MacDonald polynomials,
and, from now on, we mostly use $t$ instead of the variable $T$, these two being related by
$t=-q/T$.
Usually there are different off-shell expansions in the character basis
associated with different braid representations of the same knot (i.e. the off-shell knot polynomials are
no longer topological invariant, but instead, at least for the torus knots,
give rise to the KP tau-functions and are represented by the time dependent matrix model),
and the off-shell polynomials $X_R^{\cal K}$ are different, but the on-shell
ones all coincide and do not depend on the choice of the braid representation.

In the present paper we elaborate on the alternative suggestion of \cite{IMMMfe} and \cite{evo}:
to expand off-shell knot polynomials {\it not} in the basis of characters
$M_Q^*(A|q,t)$, as in \cite{AS,DMMSS,Che,MMMkn1,AnoMMMtree},
but in some very different basis formed by
\be\label{setI}
d[I] =
d^{i_1\ldots i_k}_{j_1\ldots j_k}
\ee
which are reduced to the
"multi-differentials" on-shell
\be
d[I]\ \to\ \prod_{a=1}^k D^{i_a}_{j_a}
\ \equiv \ \prod_{a=1}^k  \{Aq^{i_a}/t^{j_a}\},
\ \ \ \ \ \ \ \ \{x\} \equiv  x-x^{-1}
\label{muldi}
\ee
These are just (Laurent) polynomials, the word "differentials"
refers to their role in the theory of Khovanov homologies.
The reason why these quantities are originally important in knot theory
is that the specialization of representation theory at particular values of $N$
is described as vanishing of some of the elementary differentials
like $\{Aq^N\}$ and $\{A/t^N\}$, and the knot polynomials
are constructed to respect these specialization properties (see \cite{DGR,IMMMfe}
and \cite{AnoMMM21} for a detailed discussion).

Thus, we have two examples of (\ref{off}):
\be
\begin{array}{lll}
x_R^{\cal K}[I]=g_R^{\cal K}[I], & e[I]=d[I]&I\hbox{ is a set of integers}\\
&&\\
x_R^{\cal K}[I]=c_R^{\cal K}[I], & e[I]=M_I&I\hbox{ is a Young diagram}
\end{array}
\ee
and the knot polynomials in the basis of multi-differentials are labeled by the expansion coefficients
$g^{\cal K}$, which play the same role as $c^{\cal K}$ in the
character expansion. Note that the topological invariance in the differential expansion fixes its normalization
to start from unity.
In both cases, of the character and the differential expansions
the bases  ($M^*_Q$ or $D[I]$)
are {\it universal}, i.e. the same for all knots.
Dependent on the knot ${\cal K}$ are
the expansion coefficients ($c_{RQ}$ and $g_R[I]$).

The crucial advantage of the differential expansion is its power to
control the dependence of the coefficients on the representation $R$
and on the parameters $q$ and $t$.
In fact, in both cases
it is a kind of double deformation of the archetypical
relation for {\it the special polynomials} \cite{DMMSS}
(i.e. the knot polynomials at $t=q=1$),
when all the differentials reduce to just a power of $\{A\}$:
\be
{\it at}\ \ t=q=1 \ \ \ \ \ \ \ \ \ \
P^{\cal K}_R = \Big(P^{\cal K}_{_\Box}\Big)^{|R|} \ \ \ \
\Longrightarrow \ \ \ \left\{\begin{array}{c}
\sum_{Q\vdash m|R|} c_{RQ}^{\cal K} M_Q^* =
\left( \sum_{Q\vdash m} c_{_\Box Q}^{\cal K} M_Q^*\right)^{|R|} \\ \\
1+\sum_I g_R^{\cal K}[I] \{A\}^{2I} = \Big(1+\sum_I g_{_\Box}^{\cal K}[I]\{A\}^{2I}\Big)^{|R|}
\end{array}\right.
\label{facspe}
\ee
However, in the case of the differential expansion there is a {\bf miracle}:
the deformation is almost straightforward.
This miracle
was clearly demonstrated
already in \cite{IMMMfe} for the figure eight knot:
starting from the special polynomial $\sigma^{4_1}=1+\{A\}^2$,
one can define {\it a procedure}, providing not only the HOMFLY but
also superpolynomials in all symmetric and antisymmetric representation.
In \cite{AnoMMM21} this procedure was  generalized to
representation $[21]$ and some evidence was provided that it can be
further extended to {\it arbitrary} representations.
Moreover, essentially the {\it same} procedure simultaneously
provides colored knot superpolynomials for the trefoil,
starting from the $Z$-expansion of its special polynomial
$\sigma^{3_1}=1-A^2\{A\}^2$.
The task of this paper is to extend this claim, that {\bf the differential expansion opens a clear way
to tame the representation dependence of knot polynomials},
to arbitrary knots.

Namely, it looks more and more plausible that when one expands
knot polynomials in the basis $D[I]$, i.e. uses the coefficients
in front of them as the new coordinates in the space of knots,
the independent coordinates are actually the coefficients
of this expansion for the special polynomial.
It is crucial, of course that what matters is not the special
superpolynomial itself, but its appropriate $Z$-representation,
i.e. some additional structure which actually knows a lot
(perhaps, all) about the knot.
The difference may seem obscure in the above example of $4_1$,
it gets a little better seen for the trefoil, with
\be
\sigma^{3_1} = -A^4 + 2A^2 = 1-A^2\{A\}^2
\ee
and becomes quite impressive already for more complicated
2-strand knots: for example,
\be
\sigma^{[2,7]} = 4A^{6} - 3A^{8}  = 1 - (3A^2+2A^6+A^{10})\{A\}^2
+ (3A^4+2A^8)\{A\}^4 - A^6\{A\}^6
\ee
it is the vector
\be
\left.\vec g_{_\Box}^{[2,7]}\right|_{q=t=1} =
\Big[1, \ -(3A^2+2A^6+A^{10}),\ (3A^4+2A^8), \ -A^6\Big]
\label{vec27}
\ee
(not just the simple two-term expression at the l.h.s.!)
which has a potential to uniquely characterize the knot.
It is a 7-component vector
\be
\vec g_{[2]}^{[2,7]}= \vec g_{_\Box}^{[2,7]}\otimes \vec g_{_\Box}^{[2,7]}=
\Big[ 1, -2(3A^2+2A^6+A^{10}),\ 2(3A^4+2A^8) +(3A^2+A^6+A^{10})^2,\ \ldots,\ A^{12}\Big]
\ee
which can serve as a starting point for the {\it two} different
and algorithmically defined $(q,t)$-deformations,
which will provide the two superpolynomials  $P^{[2,7]}$
in representations $R=[2]$ and $R=[11]$.
(In fact, things are even more involved:
as explained in \cite{IMMMfe},
from the point of view of the $Z$-expansion the vector like (\ref{vec27})
still has an internal structure, which is signalled about by
non-unity coefficients, and the truly relevant coordinates
contain even more components, see
examples in Sects. 3-4 below.)

In other words, there is a growing evidence
that $g_{_\Box}^{\cal K}[I]$ contain considerably
more information about the knot ${\cal K}$ than $c_{_\Box Q}^{\cal K}$,
therefore, in this parametrization the deformation
of (\ref{facspe}) can be understood much better.
One can even think that it can be fully algorithmic, then the
{\bf collection of functions $g_{_\Box}^{\cal K}[I](A|q,t)$
with $R=\Box$(!), i.e. associated with the fundamental representation only}
(perhaps, even of $g_{_\Box}^{\cal K}[I](A|q=t)$, i.e. these variables
in the HOMFLY set)
{\bf could provide the {\it complete} information about the knot polynomials}.\footnote{
This conjecture looks very probable for the HOMFLY polynomials and for
the superpolynomials in the (anti)symmetric representations, while its present status for more complicated
representations remains obscure.}
This strangely sounding claim (which we illustrate below by numerous
examples) can seem to contradict the fact that the colored superpolynomials
contain much more information than the fundamental HOMFLY ones.
The secret is, of course, that the set
$g_{\Box}^{\cal K}[I]$ contains much more than just $H_{_\Box}^{\cal K}$.
The latter is obtained from the former one on-shell, i.e. from (\ref{off}) when
$d[I]$ are not just free parameters, but are substituted from
(\ref{muldi}); after that a lot of cancelations take place
and the expression can drastically simplify.
We shall see in Sect. \ref{2strtor} that the 2-strand torus fundamental HOMFLY polynomials
which are just quadratic polynomials in $A$ (up to the normalization),
have huge sets of non-zero $g_{_\Box}[I]$ quantities, which are polynomials
of high degrees in $A$.
As an opposite side of the medal, it is not at all simple to extract
$g$-variables even if the HOMFLY polynomials are known: either one should know them in
many enough representations, or possess some deep insight into the
hidden structure of differential expansions.
Anyhow, {\bf we propose that this structure does exist,
and $g$-variables provide a nice set of coordinates in the space of knots.}
They can look excessively complicated in  particular examples,
but instead {\bf they adequately reflect the structure of relations
between various knot polynomials}.
In particular, the fourth grading proposed in \cite{GGS}
may be interpreted as a kind of transform of the $g$-variables
(and there are many others of this kind, perhaps, less elegant,
but equally allowed): if this is true, this extra grading is very
different from $A,q,t$.

An intuitive picture is that with each representation $R$ there is associated
a couple of operations, which together convert
$g_{_\Box}$ variables into $g_R$:
\be
\boxed{
g_R[I_R] = \Big(g_{_\Box}[I]\Big)^{\circ |R|}
}
\ee
One of these operations, $I_R$, transforms the set $I$,
while the other one, $\circ_R$ defines an appropriate convolution
of the coefficient functions.
Still, this insight is already sufficient
to make the study of differential hierarchy of \cite{IMMMfe}
and its comparison with the character expansion, sum over paths and evolution method
quite interesting and important.

In what follows we begin with the simplest example of the figure eight knot $4_1$
in symmetric representations, where the functions $g[I]$ are
essentially trivial, hence, what remains is basically the operation $I_R$.
After that, the example of trefoil $3_1$ demonstrates that the operation
$\circ$ can also be rather simple.
However, already for the more general twist knots there are problems,
and the story requires further study.

\section{Basic examples of differential hierarchy: (anti)symmetric representations}

\subsection{Figure eight ($4_1$) knot \cite{IMMMfe,AnoMMM21}}
\label{sec:AntiSymmetricFigureEight}

Despite in the braid representation being at least $3$-strand,
the figure eight knot, the simplest of the fully symmetric knots, appears to be also the simplest
from the point of view of the colored knot polynomials and especially
of the differential hierarchy.
In particular, the answer for the
knot polynomial in the fundamental representation is just trivial:
\be
\boxed{
H_{_\Box}^{4_1}(A|q) = 1 + \{Aq\}\{A/q\}
\ \Longrightarrow \
P_{_\Box}^{4_1}(A|q,t) = 1 + \{Aq\}\{A/t\} = 1+D^{10}_{01}
}
\ee
This boxed formula encodes the first basic property of the differential hierarchy:
formulas for the HOMFLY polynomials are directly lifted to formulas for the superpolynomials,
once they are written in terms of the multi-differentials (\ref{muldi}).
In this case the only non-vanishing parameter is
\be
{\cal K} = 4_1: \ \ \ \ \ \
g^{4_1}_{_\Box}\!\!\left[^{01}_{10}\right] = 1
\ee

The second crucial feature is reflected in the archetypical formulas \cite{IMMMfe}
for the colored knot polynomials in the symmetric representations:
\be
\boxed{\
P_{[r]}^{4_1}(A|q,t) = 1 + \sum_{j=1}^r \frac{[r]_q!}{[j]_q![r-j]_q!}
\prod_{i=0}^{j-1} D^{r+i}_0D^{i}_1
\ }
\label{symm41}
\ee
Quantum numbers here are defined as $[x]_q =\frac{q^x-q^{-x}}{q-q^{-1}} = [x]_{1/q}$.
In the antisymmetric representations, the answers
\be
P_{[1^r]}^{4_1}(A|q,t) = 1 + \sum_{j=1}^r \frac{[r]_{t}!}{[j]_t![r-j]_t!}
\prod_{i=0}^{j-1} D^{0}_{r+i}D^{1}_i
\label{asymm41}
\ee
are obtained from (\ref{symm41})
by the "mirror" transform \cite{DMMSS,GS}
\be
q\longleftrightarrow -t^{-1},\ \ \ \ \ \
{\rm i.e.} \ \ \ \ \ D^i_j \longleftrightarrow D^j_i
\ee
and therefore do not seem to contribute anything new;
however, the knowledge of the both formulas is important for the
study of generic representations.
They can be rewritten in a variety of ways, one of the
most important converting the $q$-binomial coefficients
into an extended set of differentials \cite{IMMMfe}.
This explicitly explains in what sense these combinatorial factors
can be interpreted as describing the {\it set} $I(R)$ in (\ref{setI}).
For example, the right way to look at the first term in the sum
(\ref{symm41}) is to substitute it by the combination
\be
\boxed{
\l[r]_q\cdot D^r_1\ = \ \sum_{i=0}^r D^{2i}_1
}
\ \ \ \ \ \ \ \ {\rm e.g.}\ \ \ \
[2]_q\cdot\{Aq^2\}\{A/t\} = \{Aq^3\}\{A/t\} + \{Aq\}\{A/t\}
\label{decobin}
\ee
with unit coefficients.
In a similar way one can deal with all the other binomial
coefficients.
Also popular are changes of notation:
to the q-Pocchammer symbols from the $q$ factorials
and to the other sets of $A,q,t$-variables.

There are two immediate lessons from these formulas:

\begin{itemize}
\item
First, under multiplication of representations the coefficients
transform in a simple way.
The set of contributing differentials in (\ref{symm41}) and (\ref{asymm41})
is restricted, so that these formulas can be written as
\be
P_{[r]}^{4_1}(A|q,t) = 1 + \sum_{j=1}^r\ g_{r|j}\
\prod_{i=0}^{j-1} D^{r+j}_0D^{j}_1, \ \ \ \ \ \
g_{r|j} = \frac{[r]_q!}{[j]_q!\, [r-j]_q!}
\ee
and
\be
P_{[1^r]}^{4_1}(A|q,t) = 1 + \sum_{j=1}^r \ \overline{g_{r|j}}\
\prod_{i=0}^{j-1} D^{0}_{r+j}D^{1}_j, \ \ \ \ \ \
\overline{g_{r|j}} = \frac{[r]_{t}!}{[j]_t!\, [r-j]_t!}
\ee
If now one considers the generating functions
\be
\pi_{[r]}^{4_1}(z) = 1 + \sum_{j=0}^r\ g_{r|j}\,z^{2j}
\ \ \ \ \ \ \ \ \ \ \ \ {\rm and} \ \ \ \ \ \ \ \ \ \ \ \
\overline{\pi}_{[1^r]}^{4_1}(z) = 1 + \sum_{j=0}^r \ \overline{g_{r|j}}\, z^{2j}
\label{pi41}
\ee
then
\be
\pi_{[r]}^{4_1}(z) = \Big[\pi_{_\Box}^{4_1}(z)\Big]_q^r
\ \ \ \ \ \ \ \ \ \ \ \  {\rm and} \ \ \ \ \ \ \ \ \ \ \ \
\overline{\pi}_{[1^r]}^{4_1}(z) = \Big[\overline{\pi}_{_\Box}^{4_1}(z)\Big]^r_t
\label{pirels41}
\ee
where the quantum power or the q-Pocchammer symbol is
\be
\l[1+x]^r_q = 1+ \sum_{j=1}^r g_{r|j} x^j = \prod_{j=1}^r (1+q^{2j-r-1}x)
\ee
These formulas demonstrate clearly that, for different representations
$([r]$ and $[1^r]$ in this case), the both operations $I_R$ and $\circ_R$
are different. At the same time, the difference is clearly controlled by
the structure of the Young diagram in a rather simple and intuitively
appealing way.
\item
Second, modulo trivial combinatorial factors the coefficients
of expansion (\ref{symm41})
do {\it not} depend on the representation:
if one defines
\be\boxed{
g_{r|j} = \frac{[r]_q!}{[j]_q!\, [r-j]_q!} G_{j}}
\ee
then {\bf the dependence on $r$ disappears from $G$}, moreover,
\be
{\rm all} \ \ \ \ \ \ \ \ \ \ \ \
G^{4_1}_{j} = 1
\label{G41}
\ee
(for other knots they are Laurent polynomials of $A$, $q$ and $t$,
with the parameter $T=-q/t$ easily restorable from the HOMFLY case of $T=-1$).
\item
Another way to encode these relations is to write a difference equation
\cite{IMMMfe} (its relation to the quantum A-polynomial equations \cite{Apol,inds} is still
obscure, instead this form of equations is nice
to establish links with the Baxter equations associated with
$5d$ gauge theories \cite{MMeqs}):
\be
P_{[r+1]}^{4_1}(A)-P_{[r]}^{4_1}(A) = \{Aq^{2r+1}\}\{A/t\} P^{4_1}_{[r]}(qA)
\label{shieq41}
\ee
The shift $A\longrightarrow qA$ at the r.h.s. is actually responsible
for a reshuffling of the set of differentials associated with the set
of relations
\be
\Big(\l[r+1]_q\{Aq^{r+1}\}-\underline{[r]_q\{Aq^r\}} = \{Aq^{2r+1}\}\Big)
\cdot\underline{\{A/t\}}\cdot\underline{\underline{1}}, \nn \\
\Big(\l[r+1]_q\{Aq^{r+2}\}-\underline{[r-1]_q \{Aq^r\}} = [2]_q\{Aq^{2r+1}\}\Big)
\cdot\underline{\{A/t\}}
\cdot\underline{\underline{\Big([r]_q\{Aq^{r+1}\}\{Aq/t\}\Big)}}, \nn \\
\Big(\l[r+1]_q\{Aq^{r+3}\}-\underline{[r-2]_q \{Aq^r\}} = [3]_q[2]_q\{Aq^{2r+1}\}\Big)
\cdot\underline{\{A/t\}}
\cdot\underline{\underline{\Big([r]_q[r-1]_q\{Aq^{r+2}\}\{Aq^{r+1}\}\{Aq^2/t\}\{Aq/t\}\Big)}},
\nn
\ee
\vspace{-0.4cm}
\be
\ldots
\ee
where the twice-underlined differentials differ from the single-underlined ones
in the previous line by the shift $A\longrightarrow qA$.
\end{itemize}

\bigskip

{\bf The above examples actually explain what we {\it want} from the $Z$-expansion,
i.e. provide a kind of its {\it description}, at least conceptual.
It labels knots by an (infinite) set of parameters $G$, which
are independent of representation, and the $Z$-expansion provides a {\it procedure}
to reconstruct the arbitrary colored superpolynomial from the knowledge of these parameters.
Examples below are given to demonstrate that such a procedure can indeed
exist.}
Also they show that the fourth grading of \cite{GGS} acts as a trivial
rescaling in the space of $G$-parameters
(but since it does not transform the multi-differentials,
the resulting knot polynomials can, and in most cases are affected, see examples in Sect. 5).

\subsection{Trefoil}
\label{sec:AntiSymmetricTrefoil}

Our next example is the trefoil $3_1$, which is also the simplest torus knot.
This example will help us to illustrate several important issues.

\begin{itemize}
\item
While the set of relevant differentials in (anti)symmetric representations can still be
labeled as $\ r|j\ $ with $\ j=1,\ldots,r\ $ (this is the characteristic feature
of all twist knots), already for the trefoil
the representation-independent $G_{j}$ become functions of $A,q,t$:
\be
G_j^{3_1}= (-)^jA^{2j}q^{j(2j-1)}t^{j(2j-3)}
= \left(-A^2\frac{q}{t}\right)^j q^{2j(j-1)}
\label{G31}
\ee
\item
The second version of this formula demonstrates, how HOMFLY is lifted
to the superpolynomial: one should introduce $t\neq q$ in the differentials
as in (\ref{muldi}), and in the coefficients one should change
\be
\boxed{
A^2\longrightarrow A^2q/t, \ \ \ \ \ \ q^2 \longrightarrow  q^2}
\label{defoT}
\ee
i.e. the $q$-variable remains intact.
This receipt is going to work for all the twist knots.
\item
Perhaps, most important, the trefoil example allows us to illustrate
the difference between the character and differential expansions: in this
example it is still not very striking (it becomes such for more
complicated torus knots), but already quite visible.
\end{itemize}

As was already mentioned, the trefoil is a torus knot $[2,3]=[3,2]$, thus, the
character expansion is given by a straightforward $t$-deformation \cite{DMMSS}
(we remind that our superpolynomials are reduced)
of the Rosso-Jones formula \cite{RJ}.
In the fundamental representation one has
\be
P^{[2,3]}_{_\Box} = A^{3}\left(t^{-3}{\{Aq\}\over\{qt\}} - q^{3}{\{A/t\}\over\{tq\}}\right)
\ee
At the same time, the $Z$-expansion of the same quantity looks quite different:
\be
P^{3_1}_{_\Box} = 1 -\frac{A^2q}{t} \{Aq\}\{A/t\} = 1- \frac{A^2q}{t}D^{10}_{01}
\ \ \ \ \stackrel{(\ref{defoT})}{\longleftarrow} \ \ \ \
H^{3_1}_{_\Box} = 1 - A^2\{Aq\}\{A/q\}
\ee
Note that the same quantity in an unstructured form looks quite different
from the both formulas:
\be
P^{\rm trefoil}_{_\Box} = -\frac{q^2}{t^2}A^4 + A^2(q^2+t^{-2}) \ \ \ \ \ (\hbox{and}\
H^{\rm trefoil}_{_\Box} = -A^4 + A^2(q^2+q^{-2})\ )
\ee
Of course all the three formulas coincide,
\be
P^{3_1} = P^{[2,3]} = P^{\rm trefoil}
\ee
but their meaning is {\it not the same},
they belong to very different classes with very different structures
and implications.
For example, $P^{[2,3]}$ is perfectly suited
for introducing the off-shell knot polynomials in the character basis \cite{MMMkn1}.
Instead, $P^{3_1}$ is most suited for continuation to other representations:
as we explained, {\bf the purpose of the $Z$-expansion is to provide a counterpart of the
factorization property (\ref{facspe}), which can be lifted
to $t\neq q\neq 1$}.
In this particular case this is, indeed, straightforward:
\small
\be
\begin{array}{cccccc}
{\rm rep}\ R: & [1]=\Box & [2] & [3] & \ldots \\    \\
t=q=1: &  1-A^2\{A\}^2  & \Big(1-A^2\{A\}^2\Big)^2 & \Big(1-A^2\{A\}^2\Big)^3 &   \\
& & = 1 - 2A^2\{A\}^2 + A^4\{A\}^4 & = 1 - 3A^2\{A\}^2 + 3A^4\{A\}^4 - A^6\{A\}^6 &  \\  \\
t=q: & 1 - A^2\{Aq\}\{A/q\} & 1 - [2]_qA^2\{Aq^2\}\{A/q\} +
&     1 - [3]_qA^2\{Aq^3\}\{A/q\} +      & \vspace{0.2cm} \\
\vspace{0.2cm}
&& + q^2A^4\{Aq^3\}\{Aq^2\}\{A\}\{A/q\}   &  + [3]_q q^2A^4\{Aq^4\}\{Aq^3\}\{A\}\{A/q\}- &  \\
&&& - q^6A^6\{Aq^5\}\{Aq^4\}\{Aq^3\}\{Aq\}\{A\}\{A/q\}    \\ \\
t\neq q: & 1 - \frac{A^2q}{t}\{Aq\}\{A/t\} & 1 - [2]_q \left(A^2\frac qt\right) \{Aq^2\}\{A/t\} +
&     1 - [3]_q\left(A^2\frac qt\right) \{Aq^3\}\{A/t\} +      & \vspace{0.2cm} \\
&& + q^2 \left(A^2\frac qt\right)^2 \{Aq^3\}\{Aq^2\}\{Aq/t\}\{A/t\}
&  + [3]_qq^2\left(A^2\frac qt\right)^2\{Aq^4\}\{Aq^3\}\{Aq/t\}\{A/t\}- & \vspace{0.2cm} \\
&&& - q^6\left(A^2\frac qt\right)^3\{Aq^5\}\{Aq^4\}\{Aq^3\}\{Aq^2/t\}\{Aq/t\}\{A/t\}    \vspace{0.3cm} \\
\end{array}
\nn
\ee
\normalsize
This table illustrates the change of both the differentials and the $G$-coefficients
along the rows and columns.
For an alternative representation with an extended set of differentials
substituting the quantum binomial coefficients see \cite{IMMMfe,evo}.

For the generic (anti)symmetric representation one has literally the same formulas as
for the figure eight knot $4_1$:
\be
\label{eq:TefoilSymmDecomposition}
P^{3_1}_{[r]} = 1 +\sum_{j=1}^r G_{j}^{3_1}\frac{[r]_q!}{[j]_q!\, [r-j]_q!}
\prod_{i=0}^{j-1} D^{r+j}_0D^{j}_1, \nn \\
P^{3_1}_{[1^r]} = 1 +\sum_{j=1}^r \bar G_{j}^{3_1}\frac{[r]_t!}{[j]_t!\, [r-j]_t!}
\prod_{i=0}^{j-1} D^{0}_{r+j}D^{1}_j
\ee
only this time, instead of (\ref{G41}), one has (\ref{G31}):
\be
G_j^{3_1}=&(-)^jA^{2j}q^{j(2j-1)}t^{j(2j-3)}
=&\left(-\frac{A^2q}{t}\right)^j q^{2j(j-1)},
\nonumber\\
\bar G_j^{3_1}=&(-)^jA^{2j}q^{-j(2j-3)}t^{-j(2j-1)}
=&\left(-\frac{A^2q}{t}\right)^j t^{-2j(j-1)}.
\ee
Finally, relations (\ref{pirels41}) between the generating functions
remains intact:
\be
\pi_{[r]}^{3_1}(z) = \Big[\pi_{_\Box}^{3_1}(z)\Big]_q^r
\ \ \ \ \ \ \ \ \ \ \ \  {\rm and} \ \ \ \ \ \ \ \ \ \ \ \
\overline{\pi}_{[1^r]}^{3_1}(z) = \Big[\overline{\pi}_{_\Box}^{3_1}(z)\Big]^r_t
\label{pirels31}
\ee
but the generating functions themselves need to be slightly modified:
they involve  a new ingredient,
a dilatation operator $\hat\delta_q:\ \ A \longrightarrow qA$, so that
\be
\pi_{[r]}^{3_1}(z) = 1 + \sum_{j=0}^r\
G_{j}^{3_1}(A^2\hat\delta_q)\frac{[r]_q!}{[j]_q!\, [r-j]_q!}\,z^{2j}
\ \ \ \ \ \ {\rm and} \ \ \ \ \ \
\overline{\pi}_{[1^r]}^{3_1}(z) = 1 + \sum_{j=0}^r \
\bar G_{j}^{3_1}(A^2\hat\delta_{-1/t})\frac{[r]_q!}{[j]_q!\, [r-j]_q!}\, z^{2j}
\label{pi31}
\ee
In this form they remain valid for all the twist knots.

\begin{itemize}
\item
While in the case of the figure eight knot $4_1$
with all the coordinates $G=1$, the superpolynomials,
at least in all (anti)symmetric representations $R$,
were described by the $R$-dependent reshuffling of the set of
differentials $D[I]$, i.e. by what we called the operation $I_R$,
already for the trefoil this is not enough:
one should complement the same, already known, $I_R$
by the operation $\circ_R$ acting on the coefficients.
At least, for the (anti)symmetric representations it is reduced
to the sequence of box-gluing operations, $\circ_{R\rightarrow R'}$,
moreover, in this case  it is enough to consider
$\circ_r= \circ_{[r]->[r+1]}$ and $\bar\circ_r= \circ_{[1^r]}\rightarrow[1^{r+1}]$,
acting on monomials of $A^2$.
From above formulas it is clear that
\be
A^2\circ_r q^{r(r-1)}A^{2r} = q^{r(r+1)}A^{2r} & \Longrightarrow &
A^2\circ_r A^{2r} = q^{2r} A^{2r+2} \nn \\
&& A^2\overline{\circ_r} A^{2r} = t^{-2r} A^{2r+2}
\label{opsym31}
\ee
\item
The difference equation (\ref{shieq41}) for the
trefoil remains almost the same:
\be
P_{[r+1]}^{3_1}(A)-P_{[r]}^{3_1}(A) = \{Aq^{2r+1}\}\{A/t\} G_{_\Box}^{3_1}(A) P^{3_1}_{[r]}(qA)
\label{shieq31}
\ee
The only difference is the factor $G_{_\Box}^{3_1}(A)=A^2$ at the r.h.s.
(while for the figure eight knot $G_{_\Box}^{4_1}(A)=1$).
One would naturally expect a convolution operation $\circ$,
attached to it, but this equation means that in this particular case
it is fully taken into account   by the same shift $A\longrightarrow qA$,
which adequately describes the operation $I_{[r]}$ on the set of the
relevant differentials.
\end{itemize}

\subsection{Twist knots \cite{inds,evo}}
\label{sec:AntiSymmetricTwistedKnots}

The general construction which is applicable to all twist knots directly generalizes the approach described in
Sect. \ref{sec:AntiSymmetricFigureEight} and \ref{sec:AntiSymmetricTrefoil}. The main idea is to decompose the polynomials in
(anti)symmetric representations in the same set of DGR-like differentials as in (\ref{symm41}), (\ref{asymm41}), and
(\ref{eq:TefoilSymmDecomposition}). For any twist knot ${\mathcal K}$ in all (anti)symmetric representations one has
\be
P^{\mathcal K}_{[r]} = 1 +\sum_{j=1}^r G_{j}^{\mathcal K}\frac{[r]_q!}{[j]_q!\, [r-j]_q!}
\prod_{i=0}^{j-1} D^{r+j}_0D^{j}_1,\nn \\
P^{\mathcal K}_{[1^r]} = 1 +\sum_{j=1}^r \bar G_{j}^{\mathcal K}\frac{[r]_t!}{[j]_t!\, [r-j]_t!}
\prod_{i=0}^{j-1} D^{0}_{r+j}D^{1}_j,
\label{eq:TwistSymmDecomposition}
\ee
where the infinite set $G_{j}^{\mathcal K}$ depend on the knot but not on the representation.

Since the twist knots are enumerated with one parameter $k\in{\mathbb Z}\backslash0$, in the remaining part of this subsection
we refer to the corresponding $G^{\mathcal K}$ as $G^{(k)}$.

At the level of special polynomials all the coefficients $G^{(k)}_j$ are trivial and according to the rule (\ref{pirels31}) are
just simple powers of $G^{(k)}_1$
\begin{equation}
G^{(k)}_j|_{q=1}=\left(G^{(k)}_1|_{q=1}\right)^j
\end{equation}
To raise this expansion to the level of HOMFLY polynomials one has to provide a natural $q$-deformation of the multiplication of
coefficients $G^{(k)}$ in such a way that
\begin{equation}
\boxed{
G^{(k)}_j=\left(G^{(k)}_1\right)^{\circ j}.
}
\label{eq:qDeformedGMult}
\end{equation}
As opposed to the trefoil (\ref{opsym31}),(\ref{shieq31}), for the generic twist knots we do not demand the operation $\circ$
to be binary. Instead, we only claim that it is universal for all the twist knots and is $j$-linear. We define and discuss this
operation in Sect. \ref{sec:CircBilinear} and \ref{sec:CircMultilinear}.

The last transition from the HOMFLY to the superpolynomials in terms of decomposition (\ref{eq:TwistSymmDecomposition}) is
quite simple: one makes the change of variables (\ref{defoT}) in the expansion coefficients $G^{(k)}_j$
\be
A^2\longrightarrow A^2q/t, \ \ \ \ \ \ q^2 \longrightarrow  q^2,
\ee
and restore $t$ in the differentials $D^{r+j}_0\cdot D^j_1$ (correspondingly in $D^0_{r+j}\cdot D^1_j$ for the
antisymmetric representations). All nontrivial part of this procedure is included in the initial knowledge of a proper set
of differentials for the twist knots in symmetric and antisymmetric representations. We discuss the generalization
of this method beyond the symmetric/antisymmetric representations and the twist knots further in the paper.

\subsubsection{Operation $\circ_1$ for twist knots, bilinear case}
\label{sec:CircBilinear}

We suggested in (\ref{eq:qDeformedGMult}) that all the expansion coefficients $G^{\mathcal K}_j$ of a given twist knot
${\mathcal K}$ are reconstructed from $G^{\mathcal K}_1$ with some proper operation $\circ$. To illustrate this statement
we start with the simplest example of $G^{(k>0)}_2$, while the general construction is presented further in
Sect. \ref{sec:CircMultilinear}.

In case of $G^{\mathcal K}_2$, the operation $\circ$ is simply bilinear, to avoid a confusion with multilinear operation
we refer to it as $\circ_1$
\begin{equation}
G^{\mathcal K}_2=G^{\mathcal K}_1\circ_1 G^{\mathcal K}_1.
\end{equation}

The first two coefficients $G$ are
\be
G_1^{(k)} &=& A\ \frac{1-A^{2k}}{\{A\}}, \nn \\
G_2^{(k)} &=& qA^2\left(\frac{1}{\{A\}\{Aq\}} - [2]_q\frac{A^{2k}}{\{A\}\{Aq^2\}}
+ \frac{q^{4k}A^{4k}}{\{Aq\}\{Aq^2\}}\right), \nn \\
\ee

For the operation $\circ_1$ one has
\begin{equation}
\boxed{
\begin{array}{cc}
\displaystyle{\rm for}\ m\leq n &\displaystyle A^{2m} \circ_1 A^{2n} = q^{4m-2}A^{2m+2n} \\
\displaystyle{\rm for}\ m> n &\displaystyle A^{2m} \circ_1 A^{2n} = q^{4n}A^{2m+2n}\\
\end{array}
}
\end{equation}
In the both cases, the degree of $q$ depends on the {\it smaller} of the two powers.
It is often convenient to look at the multiplication table:
\begin{equation}
\label{tab:BinaryMultPos}
\begin{array}{c|c|c|c|c|c|c|cc}
\circ_1 &   A^2 & A^4 & A^6 & A^8 & A^{10} & A^{12} &\ldots & \\
\hline\hline
&&&&&&&& \\
 A^2 &\boxed{q^2}&q^2&q^2&q^2&q^2&q^2&q^2& \\
 &&&&&&&& \\
 \hline
 &&&&&&&& \\
  A^4 &q^4&\boxed{q^6}&q^{6}&q^{6}&q^{6}&q^{6}&q^6& \\
   &&&&&&&& \\
 \hline
 &&&&&&&& \\
  A^6 &q^4&q^8&\boxed{q^{10}}&q^{10}&q^{10}&q^{10}&q^{10}& \\
   &&&&&&&& \\
 \hline
 &&&&&&&& \\
  A^8 &q^4&q^8&q^{12}&\boxed{q^{14}}&q^{14}&q^{14}&q^{14}& \\
  &&&&&&&& \\
 \hline
 &&&&&&&& \\
  A^{10} &q^4&q^8&q^{12}&q^{16}&\boxed{q^{18}}&q^{18}&q^{18}& \\
   &&&&&&&& \\
 \hline
 &&&&&&&& \\
  A^{12} & q^4&q^8&q^{12}&q^{16}&q^{20}&\boxed{q^{22}}&q^{22}& \\
 &&&&&&&& \\
 \hline
 &&&&&&&& \\
 \ldots & q^4&q^8&q^{12}&q^{16}&q^{20}& q^{24} && \\
 &&&&&&&& \\
\end{array}
\end{equation}

\subsubsection{Operation $\circ$ for twist knots, multilinear case}
\label{sec:CircMultilinear}
In this section we define the multilinear operation proposed in (\ref{eq:qDeformedGMult}). This operation acts on the graded
space of polynomials in $A$ and preserves the corresponding grading, at the locus $q=1$ it turns into the simple
multiplication of polynomials in $A$. However, the operation defined here is by no means commutative or even associative.

In order to define the multilinear operation on polynomials in $A$, it is enough to describe its action on powers of $A$.
Let us have the product of $n$ monomials of $A$, which we denote by $\{k_1,k_2,\dots,k_n\}$. Since our operation preserves
the grading the result is proportional to $A^{\sum_{i=1}^n k_i}$,
\begin{equation}
\circ\left(A^{k_1},A^{k_2},A^{k_3},\dots,A^{k_n}\right)=C_{\textbf{k}} A^{\sum_{i=1}^n k_i}.
\end{equation}

The corresponding coefficient of proportionality $C_{\textbf{k}}=c^{\textrm{a}}_{\textbf{k}}c^{\textrm{f}}_{\textbf{k}}$
consists of two parts: $c^{\textrm{a}}_{\textbf{k}}$ is the average power of $q$ which depends on the set of powers, but not
on the permutation, and the so called "fine structure" $c^{\textrm{f}}_{\textbf{k}}$ which depends only on the permutation
with repetitions, but not on the exact values of powers $\textbf{k}$.

The known formulas for $G_j^{(k)}$ in the case of polynomials for the twist knots in (anti)symmetric representations uniquely
determine the first coefficient $c^{\textrm{a}}_{\textbf{k}}$ as a function of $\textbf{k}$. Whereas for the second
coefficient we could state only that the sum of $c^{\textrm{f}}_{\textbf{k}}$ over all permutations of $\textbf{k}$
with repetitions gives us a $q$-deformed multinomial coefficient. As a possible realization of such a structure, one could take
\begin{equation}
\label{eq:MultilinearCFine}
c^{\textrm{f}}_{\textbf{k}}=q^{2\#(\textrm{inversions})-\max \#(\textrm{inversions})},
\end{equation}
where $\#(inversions)$ counts inversions in the permutation with repetitions $\{k_1,k_2,\dots,k_n\}$.

The average power of $q$ is far more interesting. As it was stated, it depends only on the values of $\{k_1,\dots,k_n\}$
but not on the permutation, which can be naturally depicted as the Young diagram $\textbf{D}$ with $\sum_{i=1}^nk_i$ boxes.
Let $\{k_{i_1},k_{i_2},k_{i_3},\dots,k_{i_n}\}$ be a partially ordered set:
$k_{i_1}\geq k_{i_2}\geq k_{i_3}\geq\dots\geq k_{i_n}$, where $k_{i_r}$ is the length of the $r$-th row in the Young Diagram
$\textbf{D}$. The above partially ordered set can be parameterized with
$\{k_{j_1}^{m_1},k_{j_2}^{m_2},k_{j_3}^{m_3},\dots,k_{j_l}^{m_l}\}$, strictly ordered set
$k_{j_1}>k_{j_2}>k_{j_3}>\dots>k_{j_l}$ with multiplicities $\{m_1,m_2,\dots,m_l\}$, $n=\sum_{j=1}^l m_j$. Next, we denote the
corresponding Young diagram of the multiplicities $m_i$ with $n$ boxes as $\textbf{d}$. Our claim is that
\begin{subequations}
\label{eq:MultilinearCAvgAll}
\begin{align}
c^{\textrm{a}}_{\textbf{k}}=&q^{2\nu(\textbf{D})-n(n-1)/2-\nu(\textbf{d}^T)},\qquad
\textrm{when}\quad\forall i,\;k_i>0,
\label{eq:MultilinearCAvgPos}
\\
c^{\textrm{a}}_{\textbf{k}}=&q^{2\nu(\textbf{D})-n(n-1)/2+\nu(\textbf{d}^T)},\qquad
\textrm{when}\quad\forall i,\;k_i<0,
\label{eq:MultilinearCAvgNeg}
\end{align}
\end{subequations}
where $\nu(\textbf{D})=\sum(i-1)\textbf{D}_i$. It is worth noting that in (\ref{eq:MultilinearCAvgNeg}) all
$\textbf{D}_i$ understood as negative, to clarify this notation we present below example \ref{num:GAlgebraTwistNeg}.

The operation $\circ$ has several rather interesting properties. As a consequence of (\ref{eq:MultilinearCAvgAll}) it does not
depend on a particular value of maximal power of $A$ (minimal in the case of negative twist knots) when it has a unit
multiplicity in the set $\{k_1,\dots,k_n\}$. This can also be noted from Table (\ref{tab:BinaryMultPos}). Also, the dependence
on $\textbf{k}$ separates into three independent parts: the number of inversions, the set of powers, and the
multiplicities of powers.

Finally, we present several different examples to illustrate the definition of $\circ$.

\begin{enumerate}
\item We start from the basic example of $\circ\left(A^2,A^2,A^2\right)$, here $k_1=k_2=k_3=2$ and the corresponding diagram is
\begin{equation*}
\textbf{D}=[2,2,2]=[2^3]=\begin{array}{|c|c|}
\hline&\\
\hline&\\
\hline&\\
\hline
\end{array}\;.
\end{equation*}
In this case one has only one value of powers $k_i$ with multiplicity $3$, thus, the diagram of multiplicities
$\textbf{d}$ consists of one row with length $3$:
\begin{equation*}
\textbf{d}=[3]=\begin{array}{|c|c|c|}
\hline
&&\\
\hline
\end{array}\;,\qquad
\textbf{d}^T=[1^3]=\begin{array}{|c|}
\hline\\
\hline\\
\hline\\
\hline
\end{array}\;.
\end{equation*}
Now, using (\ref{eq:MultilinearCAvgPos})one gets
\be
\circ\left(A^2,A^2,A^2\right)=c^{\textrm{a}}_{\{2,2,2\}}A^6=q^6A^6
\ee
The above example does not illustrate (\ref{eq:MultilinearCFine}), since there is no nontrivial permutation in the set
$\{2,2,2\}$ and the number of inversions is always $0$, which means that $c^{\textrm{f}}_{\{2,2,2\}}=1$ and plays no role.
\item We continue with less trivial example consisting of different powers of $A$, namely we describe all permutations of
$\circ\left(A^2,A^2,A^6\right)$:
\begin{equation*}
\textbf{D}=[6,2,2]=[6^1,2^2]=\begin{array}{|c|c|c|c|c|c|}
\hline&&&&&\\
\hline&\\
\cline{1-2}&\\
\cline{1-2}
\end{array}\;,\qquad
\textbf{d}=\textbf{d}^T=[2,1]=\begin{array}{|c|c|}
\hline&\\
\hline\\
\cline{1-1}
\end{array}.
\end{equation*}
Again, using (\ref{eq:MultilinearCAvgPos}), one gets $c^{\textrm{a}}_{\{2,2,6\}}=c^{\textrm{a}}_{\{2,6,2\}}=
c^{\textrm{a}}_{\{6,2,2\}}=q^8$. Next, we should take into account (\ref{eq:MultilinearCFine})
\begin{equation*}
c^{\textrm{f}}_{\{2,2,6\}}=q^{-2},\qquad c^{\textrm{f}}_{\{2,6,2\}}=1,\qquad c^{\textrm{f}}_{\{6,2,2\}}=q^2.
\end{equation*}
Finally, this leads us to
\be
\circ\left(A^2,A^2,A^6\right)&=&q^6A^{10}\nonumber\\
\circ\left(A^2,A^6,A^2\right)&=&q^8A^{10}\nonumber\\
\circ\left(A^6,A^2,A^2\right)&=&q^{10}A^{10}
\ee
\item The next example includes a nontrivial multinomial coefficient when taking the sum over all permutations. To this end,
we describe all permutations of $\circ\left(A^2,A^2,A^4,A^6\right)$ which include three types of powers. Here,
\begin{equation*}
\textbf{D}=[6,4,2,2]=[6^1,4^1,2^2]=\begin{array}{|c|c|c|c|c|c|}
\hline&&&&&\\
\hline&&&\\
\cline{1-4}&\\
\cline{1-2}&\\
\cline{1-2}
\end{array}\;,\qquad
\textbf{d}=[2,1,1]=\begin{array}{|c|c|}
\hline
&\\
\hline\\
\cline{1-1}\\
\cline{1-1}
\end{array}\;,\qquad
\textbf{d}^T=[3,1]=\begin{array}{|c|c|c|}
\hline&&\\
\hline\\
\cline{1-1}
\end{array}
\end{equation*}
\begin{equation*}
c^{\textrm{a}}_{\textbf{k}}=q^{21}
\end{equation*}
\be
\circ\left(A^2,A^2,A^4,A^6\right)&=&q^{16}A^{14}\nonumber\\
\circ\left(A^2,A^2,A^6,A^4\right)&=&q^{18}A^{14}\nonumber\\
\circ\left(A^2,A^4,A^2,A^6\right)&=&q^{18}A^{14}\nonumber\\
\circ\left(A^2,A^4,A^6,A^2\right)&=&q^{20}A^{14}\nonumber\\
\circ\left(A^2,A^6,A^2,A^4\right)&=&q^{20}A^{14}\nonumber\\
\circ\left(A^4,A^2,A^2,A^6\right)&=&q^{20}A^{14}\nonumber\\
\circ\left(A^2,A^6,A^4,A^2\right)&=&q^{22}A^{14}\nonumber\\
\circ\left(A^4,A^2,A^6,A^2\right)&=&q^{22}A^{14}\nonumber\\
\circ\left(A^4,A^6,A^2,A^2\right)&=&q^{22}A^{14}\nonumber\\
\circ\left(A^4,A^6,A^2,A^2\right)&=&q^{24}A^{14}\nonumber\\
\circ\left(A^6,A^2,A^4,A^2\right)&=&q^{24}A^{14}\nonumber\\
\circ\left(A^6,A^4,A^2,A^2\right)&=&q^{26}A^{14}
\ee
\item \label{num:GAlgebraTwistNeg}
The final example illustrates formula (\ref{eq:MultilinearCAvgNeg}) for negative powers of $A$ (this corresponds to the
negative twist knots). Consider $\circ(A^{-2},A^{-2},A^{-4})$ and its permutations. Here one has
\begin{equation}
\textbf{D}=[-4,-2,-2]=[-4^1,-2^2]=
\begin{tabular}{|c|c|c|c|}
\hline
-&-&-&-\\
\hline
-&-\\
\cline{1-2}
-&-\\
\cline{1-2}
\end{tabular},
\qquad
\nu(\textbf{D})=-6,
\end{equation}
\begin{equation}
\textbf{d}=\textbf{d}^T=[2,1]=\begin{tabular}{|c|c|}
\hline
&\\
\hline
\\
\cline{1-1}
\end{tabular},
\qquad
\nu(\textbf{d}^T)=1,
\end{equation}
as a result one has
\begin{equation}
c^{\textrm{a}}_{\textbf{k}}=q^{-14}.
\end{equation}
Then, taking into account (\ref{eq:MultilinearCFine}) one gets
\begin{align}
\circ\left(A^{-4},A^{-2},A^{-2}\right)&=q^{-16},\nonumber\\
\circ\left(A^{-2},A^{-4},A^{-2}\right)&=q^{-14},\nonumber\\
\circ\left(A^{-2},A^{-2},A^{-4}\right)&=q^{-12}.
\end{align}
\end{enumerate}

\subsection{2-strand torus knots
\label{2strtor}}

This is the first case when the differential $Z$-expansion strikingly
deviates from the character expansion.
In the latter case, for all 2-strand knots one has an elementary
two-term formula for the HOMFLY polynomial in the fundamental representation
\be
H^{[2,2k+1]}_{_\Box} = \frac{A^{2k+1}}{\{q^2\}}\Big(q^{-2k-1}\{Aq\} - q^{2k+1}\{A/q\}\Big)
\ee
Instead, the $Z$-expansion is far more complicated,
resembling in complexity the alternative representation in terms of $2k+1$
strands (which also has the order $2k$ in the differentials).

Namely, we construct a formula of the form:
\be
\label{eq:ZDec2SFundamental}
H^{[2,2k+1]}_{_\Box} = 1 + \sum_{j=1} g_{_\Box|j}^{[2,2k+1]}
\prod_{i=0}^{j-1} D^{r+i}_0D^{i}_1
\ee
with the coefficients $g_{_\Box|j}$ presented in the tables below.
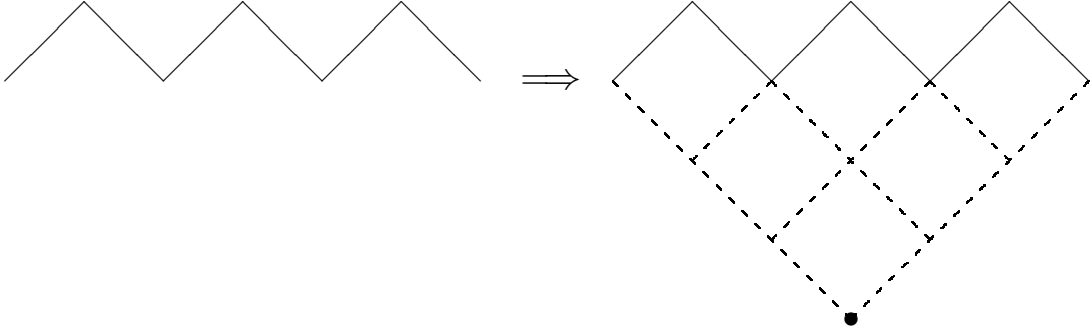
\begin{figure}[h!]
\begin{center}
\begin{picture}(410,120)(0,-90)
\put(0,0){\line(1,1){30}}
\put(30,30){\line(1,-1){30}}
\put(60,0){\line(1,1){30}}
\put(90,30){\line(1,-1){30}}
\put(120,0){\line(1,1){30}}
\put(150,30){\line(1,-1){30}}
\put(195,-3){\Large$\Longrightarrow$}
\put(230,0){\line(1,1){30}}
\put(260,30){\line(1,-1){30}}
\put(290,0){\line(1,1){30}}
\put(320,30){\line(1,-1){30}}
\put(350,0){\line(1,1){30}}
\put(380,30){\line(1,-1){30}}
\dashline{3}(230,0)(320,-90)
\dashline{3}(320,-90)(410,0)
\dashline{3}(260,-30)(290,0)
\dashline{3}(350,0)(380,-30)
\dashline{3}(290,-60)(350,0)
\dashline{3}(350,-60)(290,0)
\put(320,-90){\circle*{5}}
\end{picture}
\end{center}
\caption{The first step of decomposition of the HOMFLY polynomial for the torus knot $T^{[2,5]}_{_\Box}$.}
\label{fig:SnakeToRhombiT34}
\end{figure}

Pictorially, in the spirit of schemes in \cite{DGR},
this means that we represent a "snake"
with two ends by a sum of rhombi, with a single point added:
remarkably, under a proper (topological invariant) normalization
this added point is exactly 1: this actually seems to be true for {\it all}
knots, not only torus: {\bf the differential expansion always begins from 1}.
The rhombus in this case is exactly the first $Z$-factor $D^1_0D^0_1$. Next, we consider $g_{_\Box|0}=1$ as a reminder
with respect to $D^1_0D^0_1$, subtract it from HOMFLY polynomial and divide the remaining part by $D^1_0D^0_1$. The next
steps are less pictorial, however in each step one takes the proper reminder and continues with the quotient.
For the HOMFLY polynomial of torus knot $T^{2,5}$ in the fundamental representation, thus, one has
\begin{equation}
H^{[2,5]}_{_\Box}=1-D^1_0D^0_1\left(\frac{A^2}q[2]_q+A^6q^2-D^2_0D^1_1A^4\right),
\end{equation}
or
\begin{equation}
g^{[2,5]}_{_\Box|0}=1,\qquad
g^{[2,5]}_{_\Box|1}=-\frac{A^2}q[2]_q-A^6q^2,\qquad
g^{[2,5]}_{_\Box|1}=A^4,
\end{equation}
further in this section we assume that everywhere $g_{\dots|0}=1$.

The reason to present a simple polynomial in a quite sophisticated form is because of the relation of this decomposition
with higher (anti)symmetric representations and because it leads directly to reconstructing the superpolynomials.

Decomposition (\ref{eq:ZDec2SFundamental}) exists for all 2-strand knots in the symmetric representations and can be
straightforwardly generalized to the antisymmetric representations,
\be
P^{[2,2k+1]}_{[r]} = 1 +\sum_{j=1}^{kr} g^{[2,2k+1]}_{[r],j}
\prod_{i=0}^{j-1} D^{r+j}_0D^{j}_1,\nn \\
P^{[2,2k+1]}_{[1^r]} = 1 +\sum_{j=1}^{kr} \bar g^{[2,2k+1]}_{[1^r],j}
\prod_{i=0}^{j-1} D^{0}_{r+j}D^{1}_j,
\label{eq:DoubleStrandSymmDecomposition}
\ee

We provide a few first examples of the 2-strand torus knots in the tables below.
Note that in all presented examples the dependence of $g_i$ on $i$
is again included in the q-binomial coefficient.
In the case of fundamental representation, the general formula has the following form
\be
g^{[2,2k+1]}_{[1],j}=(-1)^j
\sum_{i=0}^{k-1}q^{j^2-jk+2ij+i}\left(\atop{k-i}{j}\right)_q\left(\atop{i+j-1}{i}\right)_q\left(\frac {A^2q}t\right)^{j+2i}
\ee

However, the dependence on the representation does not separate so nicely as in the case of the twist knots. As in
that case, it is totaly included in the combinatorial coefficients, which leads us to the point that decomposition
(\ref{eq:DoubleStrandSymmDecomposition}) is not the finest structure of the knot invariant. The corresponding coefficients
$g^{2,2k+1}_{[r],j}$ are still not the desired coordinates in the space of knot polynomials, but instead a sum of some
proper finer coordinates. The discussion and further generalizations are presented in Sect.\ref{sec:ZExpansionBeyoundSymmetric}.

In the tables below we list a few first $g_i$ for the first few 2-strand torus knots. For the sake of
brevity, we use the notation $a^2=-A^2q/t$. With this variable (first used in \cite{DGR}) all the coefficients become
positive.

\begin{table}[h!]
\begin{tabular}{c|ccc}
$T^{2,3}$&$\Box$&$\Box\Box$&$\Box\Box\Box$\\
\hline
$g_1$&$\at^2$&$\at^2[2]_q$&$\at^2[3]_q$\\
$g_2$&&$\at^4q^2$&$\at^4q^2[3]_q$\\
$g_3$&&&$\at^6q^6$
\end{tabular}
\end{table}

\begin{table}[h!]
\small
\begin{tabular}{c|lll}
$T^{2,5}$&$\Box$&$\Box\Box$&$\Box\Box\Box$\\
\hline
$g_1$&$\dfrac{\at^2}q[2]_q+\at^6q^2$&$\dfrac{\at^2}{q^2}[4]_q+\at^6q^4[2]_q$&$\dfrac{\at^2}{q^3}[6]_q+\at^6q^6[3]_q$\\
$g_2$&$\at^4$&$\frac{\at^4}{q^2}\dfrac{[3]_q[4]_q}{[2]_q}+\at^8q^6[2]_q[2]_q+\at^{12}q^{14}$&$\dfrac{\at^4}{q^4}\dfrac{[5]_q[6]_q}{[2]_q}+\at^8q^7[4]_q[3]_q+\at^{12}q^{18}[3]_q$\\
$g_3$&&$\at^6[4]_q+\at^{10}q^{10}[2]_q$&$\dfrac{\at^6}{q^3}\frac{[4]_q[5]_q[6]_q}{[2]_q[3]_q}+\at^{10}q^{10}\dfrac{[3]_q^2[4]_q}{[2]_q} +\at^{14}q^{23}[2]_q[3]_q+\at^{18}q^{36}$\\
$g_4$&&$\at^8q^4$&$\at^8\dfrac{[5]_q[6]_q}{[2]_q}+\at^{12}q^{15}[3]_q[4]_q+\at^{16}q^{30}[3]_q$\\
$g_5$&&&$\at^{10}q^5[6]_q+\at^{14}q^{22}[3]_q$\\
$g_6$&&&$\at^{12}q^{12}$
\end{tabular}
\end{table}

\begin{table}[h!]
\begin{tabular}{c|l}
$T^{2,5}$&$\Box\Box\Box\Box$\\
\hline
$g_1$&$\dfrac{\at^2}{q^4}[8]_q+\at^6q^8[4]_q$\\
$g_2$&$\dfrac{\at^4}{q^6}\dfrac{[7]_q[8]_q}{[2]_q}+\at^8q^8[4]_q[6]_q+\at^{12}q^{22}\dfrac{[3]_q[4]_q}{[2]_q}$\\
$g_3$&$\dfrac{\at^6}{q^6}\dfrac{[6]_q[7]_q[8]_q}{[2]_q[3]_q}+\at^{10}q^{10}\dfrac{[4]_q[5]_q[6]_q}{[2]_q} +\at^{14}q^{26}\dfrac{[3]_q[4]_q^2}{[2]_q}+\at^{18}q^{42}[4]_q$\\
$g_4$&$\dfrac{\at^8}{q^4}\dfrac{[5]_q[6]_q[7]_q[8]_q}{[2]_q[3]_q[4]_q}+\at^{12}q^{14}\dfrac{[4]_q^2[5]_q[6]_q}{[2]_q[3]_q} +\at^{16}q^{32}\dfrac{[3]_q^2[4]_q^2}{[2]_q^2}+\at^{20}q^{50}[2]_q[4]_q+\at^{24}q^{68}$\\
$g_5$&$\at^{10}\dfrac{[6]_q[7]_q[8]_q}{[2]_q[3]_q}+a^{14}t^{14}q^{20}\frac{[4]_q[5]_q[6]_q}{[2]_q} +\at^{18}q^{40}\dfrac{[3]_q[4]_q^2}{[2]_q}+\at^{22}q^{60}[4]_q$\\
$g_6$&$\at^{12}q^6\dfrac{[7]_q[8]_q}{[2]_q}+\at^{16}q^{28}[4]_q[6]_q+\at^{20}q^{50}\dfrac{[3]_q[4]_q}{[2]_q}$\\
$g_7$&$\at^{14}q^{14}[8]_q+\at^{18}q^{38}[4]_q$\\
$g_8$&$\at^{16}q^{24}$
\end{tabular}
\end{table}

\begin{table}[h!]
\begin{tabular}{c|ll}
$T^{2,7}$&$\Box$&$\Box\Box$\\
\hline
$g_1$&$\dfrac{\at^2}{q^2}[3]_q+\at^6q[2]_q+\at^{10}q^4$&$\dfrac{\at^2}{q^4}[6]_q+\at^6q^2[4]_q+\at^{10}q^8[2]_q$\\
$g_2$&$\dfrac{\at^4}{q^2}[3]_q+\at^8q^3[2]_q$&$\dfrac{\at^4}{q^6}\dfrac{[5]_q[6]_q}{[2]_q}+\at^8q^2[4]_q^2+\at^{12}q^{10}\left(1+[3]_q^2\right)+\at^{16}q^{18}[2]_q^2+\at^{20}q^{26}$\\
$g_3$&$\at^6$&$\dfrac{\at^6}{q^6}\dfrac{[4]_q[5]_q[6]_q}{[2]_q[3]_q}+\at^{10}q^4\dfrac{[3]_q[4]_q^2}{[2]_q}+\at^{14}q^{14}\left([2]_q+[4]_q[3]_q\right) +\at^{18}q^{24}[4]_q$\\
$g_4$&&$\dfrac{\at^8}{q^4}\dfrac{[5]_q[6]_q}{[2]_q}+\at^{12}q^8[4]_q^2+\at^{16}q^{20}\dfrac{[3]_q[4]_q}{[2]_q}$\\
$g_5$&&$\at^{10}[6]_q+a^{14}t^{14}q^{14}[4]_q$\\
$g_6$&&$\at^{12}q^6$
\end{tabular}
\label{tab:P27dec-1}
\end{table}
\begin{table}[h!]
\begin{tabular}{c|lll}
$T^{2,7}$&$\Box\Box\Box$\\
\hline
$g_1$&$\dfrac{\at^2}{q^6}[9]_q+\at^6q^3[6]_q+\at^{10}q^{12}[3]_q$\\
$g_2$&$\dfrac{\at^4}{q^{10}}\dfrac{[8]_q[9]_q}{[2]_q}+\at^8q[6]_q[7]_q+\at^{12}q^{12}\dfrac{[3]_q[4]_q[5]_q}{[2]_q}+\at^{16}q^{23}[3]_q[4]_q +\at^{20}q^{34}[3]_q$
\end{tabular}
\end{table}

\begin{table}[h!]
\small
\begin{tabular}{c|lll}
$T^{2,9}$&$\Box$&$\Box\Box$&$\Box\Box\Box$\\
\hline
$g_1$&$\dfrac{\at^2}{q^3}[4]_q+\at^6[3]_q+\at^{10}q^3[2]_q+\at^{14}q^6$&$\dfrac{\at^2}{q^6}[8]_q+\at^6[6]_q+\at^{10}q^6[4]_q+\at^{14}q^{12}[2]_q$\\
$g_2$&$\dfrac{\at^4}{q^4}\dfrac{[3]_q[4]_q}{[2]_q}+\at^8q[3]_q[2]_{q}+\at^{12}q^6[3]_q$\\
$g_3$&$\dfrac{\at^6}{q^3}[4]_q+\at^{10}q^4[3]_q$\\
$g_4$&$\at^8$
\end{tabular}
\end{table}

\bigskip

\subsection{3-strand torus knot $[3,4]$}
\label{sec:T34}

In the 3-strand case the $Z$-expansion can be obtained by a remarkable
new trick, which appears intimately related to the new grading of \cite{GGS}.

Thus, we consider the difference
\be
P^{[3,k]} - P^{[2,2k-1]} = p^{(k)},\qquad k\neq0\bmod3
\ee
and obtain its $Z$-expansion.
It turns out that while for the two-strand knots in the symmetric representation
parameters $g^{[2,2k+1]}$ are actually functions of only two variables, that is, of $q$ and of the product $\at^2\equiv -A^2q/t$,
in the deviation $p^{(k)}$, the corresponding
coefficients have one extra power of $q/t$.
Thus, a natural rescaling (related with the fourth grading from \cite{GGS})
\be
A \longrightarrow A\sigma, \ \ \ \ t \longrightarrow t \sigma^2
\ee
does not affect the coefficients $g$ for $P^{[2,2k+1]}$ and for the twist knots in (anti)symmetric representations, but does affect
those for $p^{(k)}$.

In the fundamental representation one has for the difference $p^{(k)}$ a rather simple formula
\begin{equation}
P^{[3,k]}-P^{[2,2k-1]}=D^1_0D^0_1\left(\frac{(-1)^{k+1}\at^{2k}}{(-t/q)^{\left\lfloor\frac{k-1}3\right\rfloor}} \sum_{j=1}^{\left\lfloor\frac{k-1}3\right\rfloor}\frac{[k-3j]_{qt}}{(-t/q)^{1-j}}\right),\qquad
k\neq0\bmod3,
\end{equation}
which in the simplest case of $k=4$ just gives
\begin{equation}
\label{eq:T34DifferenceFundamental}
P^{[3,4]}_{_\Box}-P^{[2,7]}_{_\Box}=\frac qt\left(\frac {A^2q}t\right)^4D^1_0D^0_1.
\end{equation}

Now, going to higher symmetric representations, one should decompose the corresponding difference $p^{(k)}$ with respect to the
same basis of multi-differentials that we used for the twist and
2-strand torus knots. This decomposition appears to be in a very intimate relation with the factorization property of
the special polynomials. At the level of special polynomials ($q=t=1$), one has for an arbitrary knot ${\mathcal K}$ and
representation $R$:
\begin{equation}
\label{eq:3StrandSpecialFactorization}
{\mathcal H}^{\mathcal K}_{R}=\left({\mathcal H}^{\mathcal K}_{_\Box}\right)^{|R|},\qquad
{\mathcal H}^{\mathcal K}_R\equiv P^{\mathcal K}_R,
\end{equation}
where $\pi_k=p^{(k)}|_{q=t=1}$. The suggestion of \cite{Anton,AntonMorozov}
was to extend (\ref{eq:3StrandSpecialFactorization})
to generic $t\neq1$ in the case of symmetric representations and to generic $q\neq1$ in the case of antisymmetric representations
\footnote{Further in this section we restrict ourselves to the symmetric representations, the antisymmetric case is absolutely
straightforward.}, i.e.
\begin{equation}
P^{\mathcal K}_{[r]}(A,q=1,t)=\left(P^{\mathcal K}_{_\Box}(A,q=1,t)\right)^r.
\end{equation}

Assume there exists a decomposition with respect to the same pairs of differentials as for the 2-strand torus and twist knots
in symmetric representations
\begin{equation}
P^{\mathcal K}_{[r]} = 1 +\sum_{j=1}^r G_{[r],j}^{\mathcal K}
\prod_{i=0}^{j-1} D^{r+j}_0D^{j}_1.
\end{equation}
Here one should note that the sequence of pairs of differentials used to decompose polynomials in the symmetric representations
turns into a simple geometric progression at $q=1$; indeed,
\begin{equation}
\left(\prod_{i=0}^{j-1}D^{r+i}_0D^i_1\right)_{q=1}=\left(\left\{A\right\}\left\{\dfrac At\right\}\right)^j,
\end{equation}
thus, $P^{\mathcal K}_{[r]}(A,q=1,t)$ can be considered as on-shell value of the generating function of the
expansion coefficients $G^{\mathcal K}_{[r],j}$ at $q=1$
\begin{equation}
\label{eq:GeneratingFunction}
f_{[r]}^{\mathcal K}(\delta)=f^{\mathcal K}_{_\Box}(\delta)=\sum_{j}\delta^jG^{\mathcal K}_{[r],j}(A,q=1,t),\qquad
\textrm{on shell:}\qquad
\delta_0=\left\{A\right\}\left\{\dfrac At\right\},\quad P^{\mathcal K}_{[r]}(A,q=1,t)=f^{\mathcal K}_{[r]}(\delta_0).
\end{equation}

Now, if one takes into account (\ref{eq:T34DifferenceFundamental}), at the locus $q=1$ one has
\begin{equation}
\label{eq:T34GeneratingFunctionDecomposition}
f_{[r]}^{[3,4]}(\delta)=\left(f_{_\Box}^{[2,7]}(\delta)+\dfrac{A^8}{t^9}\delta\right)^r=\sum_{i=0}^{r}\left(\atop ri\right)\left(f^{[2,7]}_{_\Box}(\delta)\right)^i\left(\dfrac{A^8}{t^9}\delta\right)^{r-i}
\end{equation}

The answer for the superpolynomials presents a $q$-deformation of (\ref{eq:T34GeneratingFunctionDecomposition}). The basic
idea is to describe separately the $q$-deformation of differentials and the $q$-deformation of coefficients
$G^{\mathcal K}_{[r],j}$. As for the differentials, it seems naturally to make the following substitution depending on
the representation $[r]$:
\begin{equation}
\delta^j\longrightarrow\prod_{i=0}^{j-1}D^{r+i}_0D^i_1.
\end{equation}
To describe the $q$-deformation of the multiplication rule for $G$, one can consider $P^{[3,4]}_{[2]}$ as a perturbation over
$P^{[2,7]}_{[2]}$, then one has for the difference $G^{[3,4]}_{[2]}-G^{[2,7]}_{[2]}$ Table \ref{tab:P342-P272}.

\begin{table}[h!]
\caption{q-deformation of $G^{[3,4]}_{[2],j}-G^{[2,7]}_{[2],j}$}
\begin{tabular}{|c|c|c|}
\hline
$j$&$\wt g^{3,4}_{[2],j}$&$\wt G^{3,4}_{[2],j}$\\
\hline
$j=1$&$-2 a^8 t^7$&$-a^8t^7[2]_q$\\
\hline
$j=2$&$6a^{10}t^9+4 a^{14} t^{13}+a^{16} t^{14}+2 a^{18} t^{17} $&$a^{10}t^9q[2]_q[3]_q+a^{14}t^{13}q^{10}\brc{[2]_q}^2+a^{16}t^{14}q^{12}+a^{18}t^{17}q^{19}[2]_q$\\
\hline
$j=3$&$-6 a^{12} t^{11}-4 a^{16} t^{15}$ &$-a^{12}t^{11}q^4[2]_q[3]_q-a^{16}t^{15}q^{15}\brc{[2]_q}^2 $\\
\hline
$j=4$&$2 a^{14} t^{13}$&$a^{14}t^{13}q^9[2]_q$\\
\hline
\end{tabular}
\label{tab:P342-P272}
\end{table}

\subsection{Summary: $Z$-expansion in the case of (anti)symmetric representations}
\label{sec:ZExpansionBeyoundSymmetric}
In this section we provide a summary of knowledge about the $Z$-expansion of knot polynomials in the case of symmetric and
antisymmetric representations. Our main claim concerns the basis of multi-differentials. Namely,

\begin{itemize}
\item
For all knots ${\mathcal K}$ the knot polynomials should be decomposed into $D^{r+j}_0D^j_1$ for the symmetric representations and
into $D^0_{r+j}D^1_j$ for the antisymmetric ones
\be
P^{\mathcal K}_{[r]} = 1 +\sum_{j=1}^{s_{\mathcal K}r} g^{\mathcal K}_{[r],j}
\prod_{i=0}^{j-1} D^{r+j}_0D^{j}_1,\nn \\
P^{\mathcal K}_{[1^r]} = 1 +\sum_{j=1}^{s_{\mathcal K}r} \bar g^{\mathcal K}_{[1^r],j}
\prod_{i=0}^{j-1} D^{0}_{r+j}D^{1}_j.
\label{eq:GenericSymmDecomposition}
\ee
where $s_{\mathcal K}=1$ for the twist knots, $s_{\mathcal K}=k$ for the 2-strand torus knots $T^{[2,2k+1]}$ etc.
This claim is checked for the twist knots, 2-strand torus knots and $T^{3,4}$ torus knot.
Although it was not checked directly for other knots there is a lot of evidence which supports this claim to be generic.
\item
In the case of 2-strand torus and twist knots, the expansion coefficients $g^{\mathcal K}_{[r],j}$ and
$g^{\mathcal K}_{[1^r],j}$ can be lifted from the HOMFLY to superpolynomials in a straightforward way. To this end,
one has to make the substitution in the expansion coefficients $g^{\mathcal K}$:
\begin{align}\label{sub}
A^2\rightarrow \frac {A^2q}t&,\qquad q\rightarrow q,\qquad\textrm{in case of symmetric representations},\nonumber\\
A^2\rightarrow \frac {A^2q}t&,\qquad q\rightarrow t,\qquad\textrm{in case of antisymmetric representations}.
\end{align}
Then, to generate the entire superpolynomial, one suffices to substitute these coefficients $g^{\mathcal K}$ into
(\ref{eq:GenericSymmDecomposition}). The result is not obtained from the HOMFLY polynomial by a simple change of variables,
since the differentials are $t$-deformed not in accordance with (\ref{sub}).
\item
For the twist knots the dependence of coefficients $g^{(k)}$ on the representation can be completely separated from the
dependence on the knot, that is,
\begin{align}
g^{(k)}_{[r],j}&=G_{j}^{(k)}\frac{[r]_q!}{[j]_q!\, [r-j]_q!},\nonumber\\
\bar g^{(k)}_{[1^r],j}&=\bar G_{j}^{(k)}\frac{[r]_t!}{[j]_t!\, [r-j]_t!}.
\label{eq:TwistGSeparationSummary}
\end{align}
In (\ref{eq:TwistGSeparationSummary}) all dependence on the representation is included in the binomial coefficient. Thus,
the coordinates $G_j^{(k)}$ ($\bar G_j^{(k)}$) parameterize the superpolynomials of a particular twist knot in all
(anti)symmetric representations.
\item
Indeed, all $G_j^{(k)}$ for the twist knots can be reconstructed from $G_1^{(k)}$ with a proper multilinear operation $\circ$:
\begin{equation}
G^{(k)}_j=\left(G^{(k)}_1\right)^{\circ j}
\end{equation}
described in Sect. \ref{sec:CircMultilinear}. This operation provides a natural $q$-deformation
(in the case of $\bar G$ this is the $t$-deformation) of the multiplication of polynomials in $A$. The reduction of the
above operation to simple multiplication at the locus $q=t=1$ is due to the factorization property of special polynomials.
\end{itemize}

Finally, we discuss the reason why the dependence on the (anti)symmetric representation $[r]$ ($[1^r]$) is not separated
from the coefficients $g^{\mathcal K}$ for all knots ${\mathcal K}$ as simply as for the twist knots. The point is that
the decomposition (\ref{eq:GenericSymmDecomposition}) is not the finest possible structure of the knot polynomials in the
(anti)symmetric representations.
Each pair of differentials, $D^{r+j}_0D^j_1$ (correspondingly $D_{r+j}^0D^1_j$) which we used in the decomposition
(\ref{eq:GenericSymmDecomposition}) is actually a sum of pairs of the finest level differentials
(see (\ref{decobin})). The twist knots belong to a particular case when all the expansion coefficients at the finest level are
related in the proper way to produce $D^{r+j}_0D^j_1$. This particular property guarantees the separation of variables in the
sum and thus lifts it to $g^{(k)}_j$.

We suggest that, with respect to this finer basis of pairs of differentials, all knot polynomials decomposes in the most natural
way, which means that the coefficients of decomposition does
not depend on the representation. However, this phenomenon cannot be clearly seen at the level of (anti)symmetric representations.
The particular properties of the set of differentials which we used to decompose the polynomials here do
not allow us to reconstruct the finest decomposition. The solution to this problem lays beyond the (anti)symmetric
representations.

The next step towards the proper finest basis of differentials is discussed in the next section. The elements of that basis in
an arbitrary representation $R$ are naturally associated with all subsets of the boxes of Young diagram $R$. When $R$ has
a single row or column, the sum over all subsets of the Young diagram $R$ reduces simply to the sum over number of elements
in the subset. This is the reason, why (\ref{eq:GenericSymmDecomposition}) is a sum over one index only.

\section{Generic representations}

\subsection{Generalities\label{colconj}}

Much less is yet known about the differential hierarchy for generic representations,
i.e. for the Young diagrams with more than one row or column.
The only published result so far is the recent \cite{AnoMMM21}
and no independent checks have been made since then of the conjectures,
which were formulated there.
Still some additional evidence in \cite{evo} and \cite{GGS}
seems to confirm that approach, which we
briefly formulate here.

\begin{itemize}
\item
The terms of the differential expansion are naturally graded,
by the power of $\{A\}$ in the corresponding expansion of the special
polynomial:
\be
P_R^{\cal K}(A|q=t=1) = \Big(\sigma_{_\Box}^{\cal K}(A)\Big)^{|R|}
\ee
where
\be
\sigma_{_\Box}(A) = 1 + \sum_k s_k(A)\{A\}^{2k}
\ee
and $s_k(1)\ne 0$.
Consider the expansion
\be
P_R^{\cal K}(A|q,t) = \sum_{k=0}  p_R^{(k)}(A|q,t)
\ee
where $p^{(k)}(A|q,t)$ vanishes as $\hbar^{2k}$
when $q=e^{{\rm const}\cdot\hbar}$, $t=e^{\hbar}$, $A=e^{N\hbar}$.
\item
For all knots and representations \cite{evo}
\be
p_R^{(0)} = 1
\ee
This is the {\it proper} normalization,
associated with the topological framing,
i.e. normalized in such a way that the knot polynomials
do not depend on the concrete braid realization.
Note that this normalization is different from the
group theory one, when the universal ${\cal R}$-matrix
is used in the calculations of, say, the HOMFLY polynomials (which corresponds to the vertical framing).
\item
Ideally, the term $p^{(k)}(A|q,t)$ is a $k$-linear combination
of the $Z$-factors \ $Z_{a|d}^{b|c}=\{Aq^a/t^b\}\{Aq^c/t^d\}$:
\be
p^{(k)}(A|q,t) = \sum_{|I|=k} g[I]\cdot Z^{\otimes k}[I]
\ee
with the coefficients $g[I]$ which can depend on ${\cal K}$, $R$ and $A,q,t$,
but do not need to vanish in the limit of $\hbar=0$.
\item
These coefficients $g[I]$ also exhibit some kind of
representation independence
and regularly depend on the knot ${\cal K}$
in any "natural" series of knots,
related by any kind of evolution.
\item
For the figure eight knot $4_1$ all the non-vanishing
coefficients $g[I]$ are q-binomials.
\item
In practice this is indeed so for rectangular diagrams,
at best.
In general something like the "$\epsilon^2$-terms" of \cite{AnoMMM21} (with $\epsilon=q-1/q)$)
can be needed, which do not look as "regular" as the terms with $Z$
and can even contain odd number of differentials.
Presumably the $\epsilon^2$-terms appear only for $k\geq 2$
and do not affect the $Z$-linear terms.
\item
When $A=t^N$ or $A=q^{-N}$, the knot polynomial is
reduced to the one in a smaller representation $R_{red}$,
respectively with one row or one column excluded.
This reduction respects the gradation:
terms of $p_R^{(k)}$ with a given $k$ reduce into $p_{R_{red}}^{(k)}$
with the same $k$:
\be
\left.p_R^{(k)} \ =\  p_{R_{red}}^{(k)}\right|_{A=t^N \ {\rm or} \ q^{-N}}
\label{redresp}
\ee
\item
As a non-trivial generalization of this reduction property,
the knot polynomials satisfy difference
relations as functions of representations, see Sect. \ref{eqs} below,
which also respect the gradation.
Sometime (e.g. in the case of (anti)symmetric representations \cite{IMMMfe})
they are immediately promoted to recurrent equations enough to fix the polynomials completely (with proper
initial conditions), though in general this is not yet achieved.
It is also unclear what exactly these simple and nicely looking relations
have to do with the sophisticated (but practically convenient) recursions
{\it a la} \cite{Apol,inds}, often referred to as {\it quantum ${\cal A}$-polynomials}.
\item
Extra gradings, like the one proposed in \cite{GGS}
can modify the differentials and $Z$-factors in a variety of ways,
but they do not seem to affect the coefficients $g[I]$,
which we suggest to consider as the true coordinates in the space of knots.
In this sense, it can happen that the new gradations do not provide
new knot invariants as compared to the set $\{g[I]\}$.
They can, however, be helpful to find the differential expansion {\it per se},
what, as we already saw, is not quite a trivial task even for the
(anti)symmetric representations.
\end{itemize}

In the rest of this section we illustrate to some extent
some of the items in this list.
The issue of the "fourth grading" from \cite{GGS}
will be addressed in a separate section \ref{ggsclaims}.

\subsection{$Z$-linear terms: self-consistent anzatz for the knot $4_1$
in arbitrary representation}

At the moment the self-consistent conjecture is known
for the $Z$-linear terms in arbitrary colored HOFMLY polynomials
for the figure eight knot.
For the one-hook diagrams $R$ it was already formulated
in \cite{AnoMMM21}, now we extend it to arbitrary $R$.
It serves as a starting point for all further extensions:
to terms of higher order in $Z$ and to other knots.
Self-consistency means that the $Z$-linear terms are nicely
reduced among themselves when $A=t^N$ and $A=q^{-N}$.
In accordance with our list of conjectures in Sect. \ref{colconj},
the $Z$-linear terms are free of the $\epsilon^2$-terms ($\epsilon=q-1/q$)
and are entirely made from the $Z$-factors,
which in the case of the HOMFLY polynomials are conveniently
parameterized as follows:
\be
Z_{i|j}^{(s)} = Z_{i+s|j-s} = \{Aq^{i+s}\}\{Aq^{s-j}\}
\ee
This strange notation implies that $Z_{i|j}$ is "shifted"
by $q^s$, and the shift $s$ plays an important role in the formulation
of our rule.

The rule consists of two parts.
First, with every box in the Young diagram $R$ we associate
a $Z$-factor, almost like in the hook formulas.
Second, we shift it, and the shift at the particular box depends
a little more tricky on its position in $R$.
Both parts are illustrated for $R=[10,10,10,8,7,4,4]$ in Figure \ref{fig:ZShiftsLinear}.
Let $\atop{(s)}{i,j}$ be an element of the Figure \ref{fig:ZShiftsLinear}, it corresponds to the linear term
$\propto Z_{i|j}^{(s)}=Z_{i+s|j-s}$. The index $i$ counts the number of boxes to the right of the particular box,
whereas the index $j$ counts the number of boxes down in accordance with the following rule
\begin{align}
i&=2\#(\textrm{boxes to the right})+1,\\
j&=2\#(\textrm{boxes down})+1.
\end{align}
The shift $s$ is constructed in a more tricky way. Let us split the Young diagram $Y$ into nested hooks $\Gamma_i$.
The first raw along with the first column form the first hook and so on, $Y=\bigcup_i\Gamma_i$. Fix
a particular $n$-th hook, say, with the column length $r_n$ and the row length $s_n$, create the subdiagram
$Y_n\equiv\bigcup_{i\ge n}\Gamma_i$ and complete it to the full $r_n\times s_n$-rectangular $B_n$. Then the shift $s$ for any
element of $\Gamma_n$ is the
difference of the numbers of boxes in $B_n/Y_n$ down and to the right of this element respectively.

\begin{figure}
\begin{tabular}{|c|c|c|c|c|c|c|c|c|c|}
\hline
$\atop{(0)}{19,13}$&$\atop{(0)}{17,13}$&$\atop{(0)}{15,13}$&$\atop{(0)}{13,13}$&
$\atop{(2)}{11,9}$&$\atop{(2)}{9,9}$&$\atop{(2)}{7,9}$&$\atop{(3)}{5,7}$&
$\atop{(4)}{3,5}$&$\atop{(4)}{1,5}$\\
\hline
$\atop{(0)}{19,11}$&$\atop{(0)}{17,11}$&$\atop{(0)}{15,11}$&$\atop{(0)}{13,11}$&
$\atop{(2)}{11,7}$&$\atop{(2)}{9,7}$&$\atop{(2)}{7,7}$&$\atop{(3)}{5,5}$&
$\atop{(4)}{3,3}$&$\atop{(4)}{1,3}$\\
\hline
$\atop{(0)}{19,9}$&$\atop{(0)}{17,9}$&$\atop{(0)}{15,9}$&$\atop{(0)}{13,9}$&
$\atop{(2)}{11,5}$&$\atop{(2)}{9,5}$&$\atop{(2)}{7,5}$&$\atop{(3)}{5,3}$&
$\atop{(4)}{3,1}$&$\atop{(4)}{1,1}$\\
\hline
$\atop{(-2)}{15,7}$&$\atop{(-2)}{13,7}$&$\atop{(-2)}{11,7}$&$\atop{(0)}{9,7}$&
$\atop{(2)}{7,3}$&$\atop{(2)}{5,3}$&$\atop{(2)}{3,3}$&$\atop{(3)}{1,1}$\\
\cline{1-8}
$\atop{(-3)}{13,5}$&$\atop{(-3)}{11,5}$&$\atop{(-3)}{9,5}$&$\atop{(-1)}{7,5}$&
$\atop{(0)}{5,1}$&$\atop{(0)}{3,1}$&$\atop{(0)}{1,1}$\\
\cline{1-7}
$\atop{(-6)}{7,3}$&$\atop{(-6)}{5,3}$&$\atop{(-6)}{3,3}$&$\atop{(-4)}{1,3}$\\
\cline{1-4}
$\atop{(-6)}{7,1}$&$\atop{(-6)}{5,1}$&$\atop{(-6)}{3,1}$&$\atop{(-4)}{1,1}$\\
\cline{1-4}
\end{tabular}
\caption{The linear terms in the $Z$-expansion for the diagram $R=[10,10,10,8,7,4,4]$:
each box
of the diagram contributes the $Z$-factor shown in this Figure.}
\label{fig:ZShiftsLinear}
\end{figure}

\subsection{Example of rectangular diagram: representation $[22]$}

This is an illustrative example:
it shows very clearly what kind of criteria can be used
to find the proper version of the $Z$-expansion.

Namely, one and the same HOMFLY polynomial for the trefoil
(provided, for example, by the Rosso-Jones formula)
can be expanded in a variety of ways,
of which we present three,
together with their counterparts for the figure eight knot
(for its HOMFLY polynomial see \cite{AnoMcabling} and \cite{GGS}):

$\bullet$ \ {\bf version 1}
\be
H_{[22]}^{3_1} =
1-A^2\Big(Z_{1|1}+Z_{1|3}+Z_{3|1}+Z_{3|3}\Big)
+\nn \\
+A^4Z_{2|2}D^0_0\Big(q^4D^2_0+q^{-4}D^{-2}_0+D^{4}_0+D^{-4}_0+A^{-1}\{q^2\}^2\Big)
+\nn \\
+ A^4Z_{2|2}Z_{3|3}\Big((q^2+1/q^2) -(q+1/q)^2A^2Z_{1|1}
+A^4Z_{1|1}Z_{2|2}\Big),
\label{H2231}
\ee
\be
H_{[22]}^{4_1} =
1+\Big(Z_{1|1}+Z_{1|3}+Z_{3|1}+Z_{3|3}\Big)
+Z_{2|2}D^0_0\Big(D^{2}_0+D^{-2}_0+D^4_0+D^{-4}_0\Big)
+\nn \\
+Z_{2|2}Z_{3|3}
\Big(2+ (q+1/q)^2Z_{1|1}+Z_{1|1}Z_{2|2}\Big),
\ee
In this case $H_{[22]}^{4_1}$ looks relatively nice,
but $H_{[22]}^{3_1}$ contains an $\epsilon^2$ term
$\{q^2\}^2 A^3D_0^0Z_{2|2}$.
The $Z$-linear terms are in accordance with the conjecture of
Sect. \ref{colconj}.

$\bullet$ \ {\bf version 2}
\be
H^{3_1}_{[22]} = 1 - [2]_q(q^5+q^{-5})A^2Z_{2|2}+[3]_q(q^4+q^{-4})A^4Z_{1|1}Z_{2|2}
-[2]_q^2A^6Z_{1|1}Z_{2|2}Z_{3|3} + A^8Z_{1|1}Z_{2|2}^2Z_{3|3}
= \nn \\
=1 - [2]_q[2]_{q^5}A^2Z_{2|2}+[3]_q[2]_{q^4}A^4Z_{1|1}Z_{2|2}
-[2]_q^2A^6Z_{1|1}Z_{2|2}Z_{3|3} + A^8Z_{1|1}Z_{2|2}^2Z_{3|3}
\label{H2231alt}
\ee
\be
H^{4_1}_{[22]} =
1 + [2]_q^2 \underline{(q^4-q^2-1-q^{-2}+q^{-4})}Z_{2|2}+[3]_q[2]_{q^2} Z_{1|1}Z_{2|2}
+[2]_q^2 Z_{1|1}Z_{2|2}Z_{3|3} +  Z_{1|1}Z_{2|2}^2Z_{3|3}\ \
\label{H2241alt}
\ee
These two expansions look quite nice from the point of view of selection of the
$Z$-factors, in particular, no $\epsilon^2$-terms are present.
However, the coefficients are much worse, especially in the case of $H^{4_1}_{[22]}$.
As manifestation of this, the $Z$-linear term underlined in (\ref{H2241alt})
is different from the one conjectured in Sect. \ref{colconj}.

$\bullet$ \ {\bf version 3}
\be
\boxed{
H^{3_1}_{[22]} =
1 - [2]_q^2A^2Z_{2|2}+[3]_q A^4Z_{2|2}(q^2Z_{3|1}+q^{-2}Z_{1|3})
-[2]_q^2A^6Z_{1|1}Z_{2|2}Z_{3|3} + A^8Z_{1|1}Z_{2|2}^2Z_{3|3}
}
\label{H2231alt2}
\ee
\be
\label{eq:H41-22ZDecWithCoeff}
\boxed{
H^{4_1}_{[22]} =
1 + [2]_q^2Z_{2|2}+[3]_q Z_{2|2}(Z_{3|1}+Z_{1|3})
+[2]_q^2Z_{1|1}Z_{2|2}Z_{3|3} + Z_{1|1}Z_{2|2}^2Z_{3|3}
}
\ee
This is what we think is the right $Z$-expansion:
no $\epsilon^2$-terms are present (as it should be for $[22]$,
which is a rectangular diagram),
$Z$-linear terms are in accordance with Sect. \ref{colconj}
and other coefficients are also nice.
Moreover, as expected, in the case of $4_1$ the coefficients
can be actually done unities: through the identities like
$
\Big(Z_{1|1}+Z_{1|3}+Z_{3|1}+Z_{3|3}\Big) = [2]_q^2Z_{2|2}
$ one can express (\ref{eq:H41-22ZDecWithCoeff}) as a sum over all subsets of the Young diagram $[2,2]$ of $Z$-factors.
It can be done in a few ways, one of the possible realizations of this kind is
\begin{align}
H^{4_1}_{[22]} =& 1+\Big(Z_{1|1}+Z_{1|3}+Z_{3|1}+Z_{3|3}\Big)+\nonumber\\
&+\Big(Z_{0|2} Z_{1|3}+Z_{3|1} Z_{1|3}+Z_{3|3} Z_{1|3}+Z_{2|0} Z_{3|1}+Z_{1|1} Z_{3|3}+Z_{3|1} Z_{3|3}\Big)+\nonumber\\
&+\Big(Z_{1|1} Z_{1|3} Z_{3|1}+Z_{1|1} Z_{3|3} Z_{3|1}+Z_{1|3} Z_{3|3} Z_{3|1}+Z_{1|1} Z_{1|3} Z_{3|3}\Big)+\nonumber\\
&+ Z_{1|1}Z_{2|2}^2Z_{3|3}
\label{H2241alt3}
\end{align}

\subsection{Recursion relations: emerging evidence
\label{eqs}}

One of the first basic results about the $Z$-expansion,
found already in \cite{IMMMfe}, is the set of simple recurrent relations,
like
\be
P_{[r+1]}^{4_1}(A) - P_{[r]}^{4_1}(A) = \{Aq^{2r+1}\}\left\{{A}/{t}\right\}
\cdot P_{[r]}^{4_1}(qA), \nn \\
P_{[1^{r+1}]}^{4_1}(A) - P_{[1^r]}^{4_1}(A) = \{Aq\}\left\{{A}/{t^{2r+1}}\right\}
\cdot P_{[1^r]}^{4_1}\left({A}/{t}\right)
\ee
These equations are much simpler than the conventional quantum ${\cal A}$-polynomials,
but instead they do not allow any simple reduction to $A=t^N$ and $A=q^{-N}$,
including that to the Jones polynomials (for $N=2$).

Generalization to more sophisticated representations require deeper investigation.
Still, something non-trivial is already known for the rectangular Young diagrams, namely,
for arbitrary $r_1$, $r_2$ and $k$
\be
\boxed{
P_{[r_1^k]} - P_{[r_2^k]} \sim \{Aq^{r_1+r_2}\}\{A/t^k\} = {\cal Z}_{r_1+r_2|k}
}
\ee
However, the coefficient in front of this $Z$-factor is not yet properly identified
and expressed through the superpolynomials in some other representations and
with somehow shifted arguments.
On the other hand,, this kind of proportionality seems to hold not only for the
figure eight knot $4_1$, but also for the trefoil $3_1$ and probably
for other knots.

\subsection{A comment on negative coefficients in torus superpolynomials\label{Cnc}}

Since the trefoil is a torus knot, all the HOMFLY polynomials are given by the Rosso-Jones formula.
The superpolynomials can be obtained in different ways.
For example, one can use the fact that trefoil is simultaneously a twist knot,
and thus belongs to the common series with $4_1$ (for which non-trivial
HOMFLY polynomials are obtained in \cite{AnoMcabling}) and they can be
$t$-deformed all together, {\it a la} \cite{evo}.

Another possibility is to use the superpolynomials suggested in \cite{Che}. However, there is a well-known problem related
with such obtained superpolynomials in higher representations:
they are generally no longer positive polynomials in the variables $(a,t,T)$.
At the same time, ref.\cite{GGS} contains several explicit examples of the improved (positive) trefoil superpolynomials
for the Young diagrams with several rows and columns.
The superpolynomials calculated following \cite{Che} have negative coefficients when colored not by rectangular
Young diagrams $[r^s]$. In the case of rectangular diagrams, the superpolynomials from \cite{Che} coincide with \cite{GGS}.
The only presented example in \cite{GGS} beyond the rectangular diagram is a trefoil knot in representation $[2,1]$.
Comparing this superpolynomial with that constructed in \cite{Che}, we find that they are in a rather interesting relation.
That is, once one rewrites them in the $(a,t,T)$-variables, one should take all powers of $t$ which enters monomials
with negative coefficient. Further we refer to this set of powers as $\pi_t$. Next, we apply the following simple operation
to all the monomials with powers of $t$ from $\pi_t$: we substitute coefficient $(-1)$ by $T$, and divide monomials
with positive coefficient by $T^2$. To illustrate this rule, we present the coefficient tables
(see Tab. \ref{tab:CherednikT23-21} and Tab. \ref{tab:GGST23-21}) of the two mentioned polynomials.
In Tab. \ref{tab:CherednikT23-21} and Tab. \ref{tab:GGST23-21} we omitted inessential common factors in order to compare these two
polynomials.

\begin{table}[h!]
\caption{Superpolynomial for the trefoil $3_1$ in representation $[2,1]$ constructed in ref. \cite{Che}.}
\label{tab:CherednikT23-21}
\begin{equation*}
\begin{array}{c|ccccccccccc}
&t^{-10}&t^{-8}&t^{-6}&t^{-4}&t^{-2}&t^0&t^2&t^4&t^6&t^8&t^{10}\\
\hline\\
a^6&0&0&0&0&0&T^{15}&0&0&0&0&0\\
a^4&0&0&T^8&T^{10}&T^{10}&-T^{10}+T^{12}&T^{12}&T^{14}&T^{14}&0&0\\
a^2&T^3&0&2T^5&-T^5+T^7&3T^7&-T^7+T^9&3T^9&-T^9+T^{11}&2T^{11}&0&T^{13}\\
a^0&1&0&2T^2&-T^2&2T^4&-T^4+T^6&2T^6&-T^6&2T^8&0&T^{10}
\end{array}
\end{equation*}
\end{table}

\begin{table}[h!]
\caption{Superpolynomial for the trefoil $3_1$ in representation $[2,1]$ presented in ref. \cite{GGS}.}
\label{tab:GGST23-21}
\begin{equation*}
\begin{array}{c|ccccccccccc}
&t^{-10}&t^{-8}&t^{-6}&t^{-4}&t^{-2}&t^0&t^2&t^4&t^6&t^8&t^{10}\\
\hline\\
a^6&0&0&0&0&0&T^{13}&0&0&0&0&0\\
a^4&0&0&T^8&T^8&T^{10}&T^{11}+T^{10}&T^{12}&T^{12}&T^{14}&0&0\\
a^2&T^3&0&2T^5&T^6+T^5&3T^7&T^8+T^7&3T^9&T^{10}+T^{9}&2T^{11}&0&T^{13}\\
a^0&1&0&2T^2&T^3&2T^4&T^5+T^4&2T^6&T^7&2T^8&0&T^{10}
\end{array}
\end{equation*}
\end{table}

\section{$Z$-expansion vs additional gradings
\label{ggsclaims}}

\subsection{Introducing additional gradings}

As we already explained in the text, reconstructing the $Z$-expansion from first few knot polynomials
is ambiguous. Hence, having a knot superpolynomial at hands, this is still a highly non-trivial problem to determine its
proper $Z$-expansion. This problem can be, however, simplified by introducing new
gradings which differs between different differentials. Actually, in order to reconstruct the $Z$-expansion of any colored
superpolynomial, one needs to introduce infinitely many new gradings, however, for smaller representations one needs only a few
ones.

In this section, we consider the simplest case of just one additional grading which was introduced in \cite{GGS} and
demonstrate how it emerges within the differential hierarchy. In fact, there is another argument for introducing additional
gradings: as
we already mentioned at the beginning of Sect. \ref{sec:T34}, it seems to be badly needed after I.Cherednik's observation
\cite{Che} that the colored torus superpolynomials
can fail to be {\it positive} beyond rectangular representations $[r^s]$.
A way out (if the problem exists at all, see a comment in Sect. \ref{Cnc} on how this
problem is solved in \cite{GGS})
is supposed to be that, what is obtained by this procedure is, in fact, the Euler characteristics of a
$t$-deformed complex, while there should be the corresponding Poincare polynomial of this complex, which
thus depends on an additional variable $T$. Totally this gives us a polynomial of 4 variables $(A,q,t,T)$.

We feel that the suggestion of an additional grading in \cite{GGS} is intimately related to the story of
the differential hierarchy,
but we are not sure that the structures implicitly referred to in \cite{GGS}
are exactly the same as ours.

Below we denote the additional grading variable through $\sigma$ (we explain its connection with variables of
\cite{GGS} later in this section). We claim that

\bigskip

\hspace{-0.8cm}\fbox{\parbox{17cm}{
\begin{itemize}
\item
In terms of variables $(A,q,t,\sigma)$ the fourth grading can be completely algorithmically reconstructed from the
$Z$-expansion. The building blocks of $Z$-expansion are pairs of the DGR-like differentials $D[I]$. Each pair consists of one
differential of the so-called type $X$ which scales
as $A\rightarrow A/\sigma$ and the other differential of the so-called type $Y$ which scales as $A\rightarrow A\sigma$.
At the same time, all expansion coefficients $G^{{\cal K}}_R$ remain fixed.
This conjecture is valid for all examples of quadruply-graded homology presented in \cite{GGS}.
We provide an explicit decomposition for each of that examples throughout Sect. \ref{sec:GGSFigureEight},
\ref{sec:GGSTrefoilSymmetric}, \ref{sec:GGSTrefoilBox}, and \ref{sec:GGST34}.
\item
The specific properties of the $Z$-expansion of superpolynomials of the 2-strand torus and twist knots in symmetric and
antisymmetric representations make introduction of the fourth grading $\sigma$ even more trivial. The key feature of
this decomposition in the above case is that the expansion coefficients $G^{\mathcal K}_{[r],j}$ or $G^{\mathcal K}_{[1^s],j}$
are indeed polynomials only in two instead of three variables, namely
\begin{subequations}
\begin{align}
\label{eq:GActualVarsSymmetric}
G^{\mathcal K}_{[r],j}&=G^{\mathcal K}_{[r],j}\left(\frac {A^2q}t,q\right),\\
\label{eq:GActualVarsAntiSymmetric}
G^{\mathcal K}_{[1^s],j}&=G^{\mathcal K}_{[1^s,j]}\left(\frac {A^2q}t,t\right),
\end{align}
\end{subequations}
for knot ${\mathcal K}$ being a 2-strand torus or twist knot. Next, we note that introduction of $\sigma$ in
differentials suggested in the previous item is made just with a simple change of variables inside
this differentials
\begin{subequations}
\begin{align}
\label{eq:SigmaSubstSymmetric}
A\rightarrow \frac A\sigma,\quad &t\rightarrow \frac t{\sigma^2}\quad\textrm{for symmetric representations,}\\
\label{eq:SigmaSubstAntisymmetric}
A\rightarrow A\sigma,\quad &q\rightarrow q\sigma^2\quad\textrm{for antisymmetric representations,}
\end{align}
\end{subequations}
which leaves respectively (\ref{eq:GActualVarsSymmetric}) and (\ref{eq:GActualVarsAntiSymmetric}) invariant.
This makes the quadruply-graded homology homogeneous for all
2-strand torus and twist knots in the case of symmetric and antisymmetric representation.

This result is due to specific properties of differentials used for decomposing the knot polynomials in the case of
symmetric representations along with degeneracy of the expansion coefficients. The explicit changes of variables are
presented in Sect. \ref{sec:GGSFigureEight} and \ref{sec:GGSTrefoilSymmetric}.
\item
The recursive relations discussed in Sect. \ref{eqs} can also be transferred naturally from the superpolynomials to
the quadruply-graded
homology. The introduction of the fourth grading into these relations is the same as in the differentials of knot polynomials.
We briefly illustrate this claim in Sect. \ref{sec:GGSRecursiveRelations}.
\end{itemize}}}

\bigskip

In fact, particular comments of this kind were already made
in appropriate places of the previous sections, now it is time
to make them a little more systematic.
In the rest of this section we explain what we mean in a little more detail.

\subsection{Figure eight knot, representations $[1]$ and $[2]$}
\label{sec:GGSFigureEight}

Both presented examples in sec.4.2 of \cite{GGS} are indeed polynomials in three variables. If one takes the
superpolynomial in the MacDonald variables $A,q,t$, then using the substitution
\be
A = \alpha\sqrt{-t_r^3t_c},\ \ \ \  q = -\kappa t_c, \ \ \ \ t = \kappa t_r
\label{chavs42}
\ee
one gets the quadruply graded answer in the variables
$(\alpha,\kappa,t_c,t_r)$.\footnote{Throughout this section we denote four variables $(a,q,t_c,t_r)$ used in \cite{GGS} as
$(\alpha,\kappa,t_c,t_r)$ to avoid a confusion with $a$ and $q$ used here.}
Or, after substitution
\be
\alpha = t_r^{-2}A\sqrt{t/q}, \ \ \ \ \kappa = t_r^{-1}t,\ \ \ \ t_c=-t_rq/t
\label{chavas42}
\ee
one gets
\be
\boxed{
{\cal P}_{[1]}^{4_1}(\alpha,\kappa,t_c,t_r) = P^{4_1}_{[1]}(A,q,t) =
1 + \{Aq\}\{A/t\}}
\ee
i.e. $t_r$ drops out of the answer.
It was already  noted in \cite{GGS}
that, in this case, the answer depends only on the product $t_rt_c$,
thus, the number of independent variables is three, not four.
But in general eliminating the fourth variable is more
sophisticated, provided by the change of variables (\ref{chavas42}).

In particular, after this substitution the fourth variable $t_r$
also drops out of
\be
{\cal P}_{[2]}^{4_1}(\alpha,\kappa,t_c,t_r) =
P^{4_1}_{[2]}(A,q,t) =
1 + (q+q^{-1})\{Aq^2\}\{A/t\} + \{Aq^3\}\{Aq^2\}\{Aq/t\}\{A/t\}
\ee

In this case, this can also be seen from the very beginning,
because the answer is homogeneous:
all the terms have the same value
\be
\label{eq:41-SymGGSInvariant}
2\#(\alpha)+\#(\kappa)-\#(t_c)-\#(t_r)=\textrm{const}=0
\ee

\subsection{Trefoil in representations $[1]$, $[2]$ and $[11]$}
\label{sec:GGSTrefoilSymmetric}

First, we note that all answers for the (anti)symmetric representations presented in sec. 4.1 of \cite{GGS} are
homogeneous and, thus, the fourth grading can be completely eliminated from them. For representations $[1]$ and
$[2]$ one has
\begin{subequations}
\begin{equation}
\label{eq:31-SymGGSInvariant}
2\#(\alpha)+\#(\kappa)-\#(t_c)-\#(t_r)=\textrm{const},
\end{equation}
with exactly the same invariant (\ref{eq:41-SymGGSInvariant}) as for the figure eight knot (the nonzero value of
$\textrm{const}$ corresponds to an inessential common factor).
At the same time, for representation $[11]$ one has another invariant, which looks completely different
\begin{equation}
\label{eq:31-11GGSInvariant}
4\#(\alpha)+\#(\kappa)-5\#(t_c)+\#(t_r)=\textrm{const}.
\end{equation}
\label{eq:31-AntiSymmetricGGSInvariantBlock}
\end{subequations}

It is worth noting that (\ref{eq:31-11GGSInvariant}) is not applicable to the trefoil in representation $[1]$, the reason
is a rather specific choice of grading $(\alpha,\kappa,t_c,t_r)$ which heavily depends on the number of rows in the
Young diagram. Additionally, the choice of grading $(\alpha,\kappa,t_c,t_r)$ makes
mirror symmetry transformation formulas dependent on the diagram (see Sect. 3.3 of \cite{GGS} for details). We prefer to use
the set of variables which makes description of the mirror symmetry universal for all diagrams:
\begin{equation}
\boxed{
\label{eq:VariablesFirstMention}
\alpha=A\sqrt{\dfrac tq},\qquad
\kappa=t\sigma^{-1/l(R)},\qquad
t_c=-\dfrac{q}t\sigma^{1/l(R)},\qquad
t_r=\sigma^{-1/l(R)},}
\end{equation}
where $l(R)$ is the number of rows in corresponding Young diagram $R$.

The dependence on the diagram in (\ref{eq:VariablesFirstMention}) compensates the corresponding dependence of the mirror
symmetry rules for the variables $(\alpha,\kappa,t_c,t_r)$. In terms of the variables $(A,q,t,\sigma)$, the mirror
transformation is the simple exchange
\begin{equation}
\boxed{
A\leftrightarrow A,\qquad
q\leftrightarrow -\dfrac1t,\qquad
\sigma\leftrightarrow\dfrac1\sigma}
\end{equation}
for all knots and representations.

Return to the regular triple-graded homology (superpolynomials) is thus achieved at $\sigma=1$. At the same time,
$A,q,t$ are just the MacDonald variables. The fourth grading $\sigma$ here is exactly the grading $Q$ in \cite{GGS}
(formulas of Sect. 1.5 and the next to last one in Sect. 2.4).

Now, if one rewrites the $Z$-expansion of the quadruply-graded constructions in terms of variables
(\ref{eq:VariablesFirstMention}),
one can note that the fourth sigma-grading can be reconstructed \textbf{fully algorithmically} from the proper
expansion of the superpolynomials, that is,

\begin{equation}
\left(\frac qt\right){\mathcal P}^{3_1}_{[1]}=1-\left(\frac {A^2q}t\right)\left\{\frac{A\sigma}t\right\}\left\{\frac{Aq}{\sigma}\right\}
\end{equation}

\begin{equation}
\left(\frac qt\right)^2{\mathcal P}^{3_1}_{[2]}=1
-\left(\frac {A^2q}t\right)\left(\left\{\frac{Aq^3}{\sigma}\right\}\left\{\frac{A\sigma}t\right\}+\left\{\frac{Aq}{\sigma}\right\}\left\{\frac{A\sigma}{t}\right\}\right)
+\left(\frac {A^2q}t\right)^2q^2\left\{\frac{Aq^3}{\sigma}\right\}\left\{\frac{Aq^2}{\sigma}\right\}\left\{\frac{Aq\sigma}t\right\}\left\{\frac{A\sigma}t\right\}
\end{equation}

\begin{equation}
\left(\frac qt\right)^2{\mathcal P}^{3_1}_{[11]}=1
-\left(\frac {A^2q}t\right)\left(
\left\{\frac{A\sigma}{t^3}\right\}\left\{\frac{Aq}{\sigma}\right\}+
\left\{\frac{A\sigma}t\right\}\left\{\frac{Aq}{\sigma}\right\}
\right)
+\left(\frac {A^2q}t\right)^2t^{-2}\left\{\frac{A\sigma}{t^3}\right\}\left\{\frac{A\sigma}{t^2}\right\}
\left\{\frac{Aq}{\sigma}\right\}\left\{\frac{Aq}{t\sigma}\right\}
\end{equation}
The $Z$-expansion building blocks are pairs made of two different types of differentials. To make the description more pictorial
one can associate one differential with the $x$-coordinate of a particular box of the Young diagram and the other one
with the $y$-coordinate (however, this does
not implies the exact separation, see Sect. \ref{sec:ZExpansionBeyoundSymmetric} for a detailed description). To restore
$\sigma$, one should scale $A$ in the first type of differentials as $A\rightarrow A/\sigma$, whereas in the second sort of
differentials one substitutes $A\rightarrow A\sigma$. The same rule also holds for the figure eight knot presented examples,
for the trefoil in representation $[22]$, and for knot $T^{3,4}$ in representations $[1],[2]$, thus for all examples
presented in \cite{GGS}.
We conjecture this to be a generic rule for reconstruction of the quadruply graded homology.

Finally, we should make a note on the specifics of the symmetric and antisymmetric representations. In terms of the variables
$(A,q,t,\sigma)$ (\ref{eq:VariablesFirstMention}), formulas (\ref{eq:31-AntiSymmetricGGSInvariantBlock}) turn into
\begin{subequations}
\begin{align}
\label{eq:HomogenetyMacDonald}
\#(A)+2\#(t)+\#(\sigma)&=const,\\
\#(A)-2\#(q)-\#(\sigma)&=const.
\end{align}
\end{subequations}
In other words, if one takes the superpolynomial of knot $3_1$ in the (anti)symmetric representation $P^{3_1}_{[r]}$
(correspondingly $P^{3_1}_{[1^r]}$) then {\bf the quadruply-graded homology can be reconstructed via the simple
change of variables}
\begin{subequations}
\label{eq:31-SymmAsymmSubstitution}
\begin{align}
\label{eq:31-SymmetricSubstitution}
{\mathcal P}^{3_1}_{[r]}&=P^{3_1}_{[r]}\left(A/\sigma,q,t/\sigma^2\right),\\
{\mathcal P}^{3_1}_{[1^r]}&=P^{3_1}_{[1^r]}\left(A\sigma,q\sigma^2,t\right).
\end{align}
\end{subequations}
{\bf We conjecture that the same should hold for all the 2-strand torus and twist knots in all symmetric and antisymmetric
representations.}

\subsection{Trefoil in representation $[22]$}
\label{sec:GGSTrefoilBox}

For representation $[22]$, after two minor misprints corrected,
the answer of \cite{GGS} decomposes nicely in terms of multi-differentials. Namely, rewrite the proposed polynomials in terms of
the variables $(A,q,t,\sigma)$ (\ref{eq:VariablesFirstMention}) for the number of rows $R=2$
\be
\alpha=A\sqrt{\frac tq},\qquad
\kappa=\frac t{\sqrt{\sigma}},\qquad
t_c=-\frac{q\sqrt{\sigma}}t,\qquad
t_r=\frac1{\sqrt{\sigma}},\qquad
(\textrm{in these terms}\quad
\gamma = t_ct_r = -q/t=T).
\label{chava22}
\ee
Then one has
\be
\left(\frac{q}{t}\right)^4 {\cal P}_{[22]}^{3_1,GGS}&=&
1-A^2\left\{\frac{Aq^2}{\sigma}\right\}\left\{\frac{A\sigma}{t^2}\right\}\times\nn\\
&&\times
\Big(-\left(\frac{q}{t}\right)^8A^6
\left\{\frac{Aq^3}{\sigma}\right\}\left\{\frac{A\sigma}{t^3}\right\}\
\left\{\frac{Aq^3}{t\sigma}\right\}\left\{\frac{Aq\sigma}{t^3}\right\}\
\left\{\frac{Aq^2}{t\sigma}\right\}\left\{\frac{Aq\sigma}{t^2}\right\} + \nn \\
&&+A^4\left(\frac{q^6}{t^4}+\frac{q^4}{t^4}+\frac{q^6}{t^6}+\frac{q^4}{t^6}\right)
\left\{\frac{Aq^3}{\sigma}\right\}\left\{\frac{A\sigma}{t^3}\right\}\
\left\{\frac{Aq^2}{t\sigma}\right\}\left\{\frac{Aq\sigma}{t^2}\right\} -\nn \\
&&-\left(\frac{q}{t}\right)^3A^2\left(qt^2(t+t^{-1})
\left\{\frac{Aq^3}{\sigma}\right\}\left\{\frac{Aq\sigma}{t^2}\right\}
+\frac{(q+q^{-1})}{q^2t}
\left\{\frac{Aq^2}{t\sigma}\right\}\left\{\frac{A\sigma}{t^3}\right\}
+\right.\nn \\
&&\left.+\left\{\frac{Aq^2t}{\sigma}\right\}\left\{\frac{Aq\sigma}{t^2}\right\}
+\left\{\frac{Aq^2}{t\sigma}\right\}\left\{\frac{A\sigma}{qt^2}\right\}
\right)
+\frac{q}{t}(q+q^{-1})(t+t^{-1})\Big)
\ee

\subsection{Three-strand torus knot $T^{3,4}$}
\label{sec:GGST34}

The only example of a three-strand torus knot in \cite{GGS} is $T^{3,4}$. Applying the change of variables
(\ref{eq:VariablesFirstMention}) to results presented in sec.4.3 of \cite{GGS} one can note that
(\ref{eq:HomogenetyMacDonald}) is no longer true. Instead, the quadruply-graded construction separates into several
homogeneous pieces. In the simplest example of the fundamental representation there are two pieces
\be
\boxed{
\left(\frac qt\right)^3{\cal P}_{[1]}^{[3,4]}(A,q,t,\sigma) =
{\Pr}_0\Big({\mathcal P}^{[3,4]}_{_\Box}\Big)
+ \dfrac qt\cdot {\Pr}_{1}\Big({\mathcal P}^{[3,4]}_{_\Box}\Big)
\label{eq:T34-1QuadGradedDecomposition}
}
\ee
where ${\Pr}_k$ denotes the projector onto the degree $k$ homogeneous part and ($P^{[2,7]}_{_\Box}$ here can be
taken from Sect. 3.4, we need it only as a function of the three variables without any references to its $Z$-expansion)
\begin{subequations}
\be
\label{eq:P34-1-deg0}
{\Pr}_0\Big({\mathcal P}^{[3,4]}_{_\Box}\Big) =& {\mathcal P}^{[2,7]}_{_\Box}&=P^{[2,7]}_{_\Box}\left(A/\sigma,q,t/\sigma^2\right), \\
\label{eq:P34-1-deg1}
{\Pr}_{1}\Big({\mathcal P}^{[3,4]}_{_\Box}\Big) =&\dfrac tq\left({\mathcal P}^{[3,4]}_{_\Box} - {\mathcal P}^{[2,7]}_{_\Box}\right)
&=\left(A^2\frac qt\right)^4
\left\{\frac{Aq}\sigma\right\}\left\{\frac{A\sigma}t\right\},
\ee
\end{subequations}
The homogeneity is understood here w.r.t. the scaling (\ref{eq:SigmaSubstSymmetric}) which allows one to reduce any homogeneous
piece to a functions of less number of variables, i.e. to remove the fourth grading $\sigma$. On contrary, in the sum of
a few pieces of different homogeneities this scaling would produce factors of $\sigma$ of different degrees.
In particular, in rbis concrete case the reason for ${\mathcal P}^{[3,4]}_{[1]}$ to be no longer homogeneous is the presence
of an additional factor of $q/t$ in the
$Z$-decomposition of this superpolynomial. Unlike the combination $A^2q/t$ the factor $q/t$ is not invariant under change
(\ref{eq:31-SymmetricSubstitution}). This reveals the origin of the fourth grading $\sigma$ as a different rescaling of two
types of differentials which enter the $Z$-decomposition in pairs. Again, we conjecture that transition to the
quadruply-graded construction does
not affect the expansion coefficients $g^{\mathcal}_{R,j}$ for all knots ${\mathcal K}$ and representations $R$.

The decomposition (\ref{eq:T34-1QuadGradedDecomposition}) is still not very impressive, due to the extreme simplicity of
superpolynomial of the torus knot $T^{3,4}$ in the fundamental representation, however, we started from this trivial example to
demonstrate the generic concept. The approach becomes more spectacular when being applied to the quadruply-graded
${\mathcal P}^{3,4}_{[2]}$. Here one has
\begin{equation}
\boxed{
\left(\frac qt\right)^6{\mathcal P}^{[3,4]}_{[2]}=\Pr_0\left({\mathcal P}^{[3,4]}_{[2]}\right)+
\frac qt\Pr_1\left({\mathcal P}^{[3,4]}_{[2]}\right)+\left(\frac qt\right)^2\Pr_2\left( {\mathcal P}^{[3,4]}_{[2]}\right),
}
\end{equation}
where
\begin{subequations}
\be
\label{eq:P34-2-deg0}
{\Pr}_0\left({\mathcal P}^{[3,4]}_{[2]}\right)=&{\Pr}_0\left(P^{[3,4]}_{[2]}\right)\left(A/\sigma,q,t/\sigma\right)&=P^{[2,7]}_{[2]}\left(A/\sigma,q,t/\sigma^2\right),\\
\label{eq:P34-2-deg1}
{\Pr}_1\left({\mathcal P}^{[3,4]}_{[2]}\right)=&{\Pr}_1\left(P^{[3,4]}_{[2]}\right)\left(A/\sigma,q,t/\sigma\right),\\
\label{eq:P34-2-deg2}
{\Pr}_2\left({\mathcal P}^{[3,4]}_{[2]}\right)=&{\Pr}_2\left(P^{[3,4]}_{[2]}\right)\left(A/\sigma,q,t/\sigma\right)&=\left(\dfrac {A^2q}t\right)^8q^{12}\left\{\dfrac {Aq^3}\sigma\right\}\left\{\dfrac {Aq^2}\sigma\right\}\left\{\dfrac{Aq\sigma}t\right\}\left\{\dfrac{A\sigma}t\right\}.
\ee
\label{eq:P34-2-degAll}
\end{subequations}
Equation (\ref{eq:P34-2-deg0}) looks like a straightforward generalization of (\ref{eq:P34-1-deg0}). As for the highest homogeneity
degree, if one compares (\ref{eq:P34-2-deg2}) with (\ref{eq:P34-1-deg1}) it is natural to treat the expansion coefficient
in the second symmetric representation as a square of the corresponding coefficient in the fundamental representation
with respect to some graded operation $\circ'$ similar to that in Sect. 3.3.1:
\begin{equation}
\at^8\circ'\at^8=q^{12}\at^{16}
\end{equation}
The most interesting part of (\ref{eq:P34-2-degAll}) is the omitted left part of equation (\ref{eq:P34-2-deg1}). Using
$\at$ as a natural variable in the expansion coefficients, one can rewrite
\be
{\Pr}_1\left({\mathcal P}^{[3,4]}_{[2]}\right)= [2]_q \left\{\dfrac{Aq^2}\sigma\right\}\left\{\dfrac{A\sigma}t\right\}\times
\nonumber\\ \times\left(\at^8-\left\{\dfrac{Aq^3}\sigma\right\}\left\{\dfrac{Aq\sigma}t\right\} \left(\left(\at^{18}q^{19}+\at^{14}q^{10}[2]_q+\at^{10}q[3]_q\right)- \left\{\dfrac{Aq^4}\sigma\right\}\left\{\dfrac{Aq^2\sigma}t\right\}\times \right.\right.
\nonumber\\
\left.\left.\times\left(\at^{16}q^{15}[2]_q+\at^{12}q^4[3]_q\right)- \left\{\dfrac{Aq^5}\sigma\right\}\left\{\dfrac{Aq^3\sigma}t\right\}\at^{14}q^9 \right)\right),
\ee
or, in terms of the coefficients $g$, one has
\begin{equation}
g=[2]_q\left(\begin{array}{c}
0\\
-\at^8\\
\at^{18}q^{19}+\at^{14}q^{10}[2]_q+\at^{10}q[3]_q\\
-\at^{16}q^{15}[2]_q-\at^{12}q^4[3]_q\\
\at^7q^9
\end{array}\right)
\end{equation}
Again, we note that in all three parts of (\ref{eq:P34-2-degAll}) the fourth grading do not affect the coefficients $g$ of decomposition in the basis of multi-differentials.

\subsection{Recursive relations}
\label{sec:GGSRecursiveRelations}

An algorithmic way of introducing the additional grading allows one to deform immediately various relations to the
quadruply-graded case.
For instance, according to the described rules, the
equations from Sect. \ref{eqs}
are immediately generalized to the case of quadruply-graded polynomials.
For example, for the trefoil
\begin{align}
\left(\frac qt\right)^2{\cal P}_{[22]} - {\cal P}_{[11]} =& \left\{\frac{Aq^3}{\sigma}\right\}
\left\{\frac{A\sigma}{t^2}\right\}\left(
A^{14} q^{15}t^{-16}-A^{12} q^{15}\sigma ^{-2} t^{-10}-A^{12} q^{13}\sigma ^{-2}t^{-12}-A^{12} q^{13}\sigma ^{-2} t^{10}-\right.\nonumber\\
&-A^{12} q^{11} \sigma ^2t^{-16}-A^{12} q^{11} \sigma ^2t^{-14}-A^{12} q^9 \sigma ^2t^{-14}+A^{10} q^{13}\sigma ^{-4} t^{-6}+A^{10} q^{13}\sigma ^{-4} t^{-4}+\nonumber\\
&+A^{10} q^{11}t^{-10}+A^{10} q^{11}t^{-8}+A^{10} q^{11}\sigma ^{-4} t^{-6}+A^{10} q^9t^{-12}+A^{10} q^9t^{-10}+A^{10} q^9t^{-8}+\nonumber\\
&+A^{10} q^7 \sigma ^4t^{-14}+A^{10} q^5 \sigma ^4t^{-14}+A^{10} q^5 \sigma ^4t^{-12}-A^8 q^{11}\sigma ^{-6}-A^8 q^9\sigma ^{-2} t^{-6}-\nonumber\\
&-A^8 q^9\sigma ^{-2} t^{-4}-A^8 q^7 \sigma ^2t^{-8}+A^8 q^5\sigma ^{-2} t^{-6}-A^8 q \sigma ^6t^{-12}-A^6 q^5\sigma ^{-4}-A^6 q^3t^{-6}-\nonumber\\
&\left.-A^6 q^3t^{-4}-A^6 q \sigma ^4t^{-8}
\right)
\end{align}
and the deformation of the differentials at the r.h.s. is as expected.

\section{Conclusion}

Of crucial importance in the study of every particular
model of quantum field/string theory
is understanding of what is  appropriate basis for
its correlation functions.
In Chern-Simons theory at least two such bases
are solidly identified: that of the chord diagrams,
relevant for the theory of Vassiliev invariants \cite{Vass}, and for genus expansion of \cite{3refs},
and that of the $SU(\infty)$ characters
(the Schur and MacDonald functions), naturally appearing
\cite{DMMSS,MMSS,MMS} in the braid realization of knots and allowing one to introduce the off-shell (extended)
knot polynomials {\it a la} \cite{MMMknI}.

In this paper we claim that the  basis, provided
by the $Z$-expansion of \cite{IMMMfe} can be of no less
importance, and for some purposes even better than
the character expansion.
There is a whole number of motivations for this study.

\begin{itemize}
\item
The story starts from the factorization property
(\ref{facspe})
of special polynomials (i.e. at $q=t=1$)
and the first purpose is to lift it
to the HOMFLY and superpolynomials as straightforwardly
as only possible.
An important sign that this is a well motivated task, was
a partial (in only one variable), but {\it literal} extension,
at least, for particular representations,
in \cite{Anton}.
\item
The second crucial observation is that the special polynomials,
functions of $A$ only are in fact naturally
expanded in powers of $A$ itself and of $\{A\}^2=(A-1/A)^2$.
This sounds strange, and of course this bi-expansion
is not defined entirely at the level of special polynomials.
Instead, it keeps some non-trivial information about structure of the
generic colored superpolynomial.
If there was $t \neq 1$, then one could put $A = t^N$
and consider an expansion in powers of
$\bar\hbar=\log t$, where $\{A\}$ would be of the order
$\bar\hbar$. The $\{A\}$-expansion is a remnant of that expansion,
and the powers of $A$ are introduced so that no new powers of $\bar\hbar$ are added.
This hidden structure information about the knot polynomial
is dramatically extended at the next step of our reasoning.
\item
This next observation is just that each $\{A\}$
is the $q=t=1$ limit of some DGR-differential
$D^i_j=\{Aq^i/t^j\}$. Thus, the $\{A\}$-expansion
with $A$-dependent coefficients of
the special polynomial
comes from the corresponding expansion of the entire
superpolynomial.
In the simplest, (anti)symmetric representations
this expansion has a peculiar form
\be
P_{[r]} =
1 + Z_{1|1}\left(g_1 + Z_{2|1}\left(g_2 + Z_{3|1}\Big(g_3
+ \ldots \Big)\right)\right)
\ee
with $Z_{i|1} =\{Aq^i\}\{A/t\}$, hence, the two names:
$Z$-expansion and differential hierarchy.
In general this expansion actually describes an arbitrary
colored superpolynomial in terms of the coefficients
$g_R[I]$ of its expansion in multi-differentials.
\item
The crucial property of this expansion
is that the set of coefficients $g_R[I](A,q,t)$ is much
simpler than it seems.
It looks like the knowledge of just $g_{_\Box}(A,q=t)$
for $R=\Box$ and for the HOMFLY polynomials $q=t$ may be finally
sufficient to find all $g_R^{[I]}$,
as certain powers of $g_{_\Box}$ w.r.t. some $R$-dependent
non-associative multiplications and comultiplications,
and with an algorithmically defined $t$-deformation.
Moreover, it looks like the other gradings, including the
one suggested in \cite{GGS}, can also be algorithmically
introduced, once the $Z$-expansion is known.
\end{itemize}

In this paper we gave only some very limited evidence
in support of these observations.
The differential hierarchy is rather tedious to work out
even in the simplest examples.
Moreover, it is not always reduced to the $Z$-factors:
as it is known since the $[21]$ example of \cite{AnoMMM21},
some "$\epsilon^2$"-terms, even with odd numbers
of differentials, can occur, which still need to be
appropriately understood and tamed.
Still, we believe that the existing evidence is already
convincing enough to justify the need to study
the differential hierarchy along with other generic
approaches to knot polynomials.

\section*{Acknowledgements}

Our work is partly supported by Ministry of Education and Science of
the Russian Federation under contract 8498, by the Brazil National Counsel
of Scientific and Technological Development (A.Mor.), by the President fund
MK-1646.2011.1 (S.A.), by
NSh-3349.2012.2, by RFBR grants RFBR 12-02-00594 (S.A.), 13-02-00457 (A.Mir.) and 13-02-00478
(A.Mor.), by 12-01-33071 mol-a-ved (S.A.), by joint grants 12-02-92108-Yaf-a,
13-02-91371-ST-a.


\begin{thebibliography}{12}
\bibitem{AntonMorozov} This conjecture for the torus knot superpolynomials only was first proposed in \cite{Che}
(for any box Young diagram). For more general knots in generic symmetric and antisymmetric representations it was
first discussed in \cite{Anton}.
\end{thebibliography}
\end{document}